\documentclass[prl,reprint,showpacs,  superscriptaddress]{revtex4-2}
\usepackage{graphicx}% Include figure files
\usepackage{amssymb}
\usepackage{amsmath}
\usepackage{amsthm}
\usepackage{newtxtext,newtxmath}
\usepackage{dsfont}
\usepackage{wasysym}
\usepackage{color}
\usepackage{array}
\usepackage{multirow} 
\usepackage{makecell}
\usepackage{xcolor}
\usepackage[colorlinks=true,linkcolor=blue,urlcolor=blue,citecolor=blue,pdfauthor={ },pdftitle={},pdfsubject={ },pdfkeywords={ }]{hyperref}
\graphicspath{{Figures/}}

\usepackage{changes}
\setaddedmarkup{\color{red} {#1}}
\setdeletedmarkup{\color{blue} \sout{#1}}
\sethighlightmarkup{{\color{orange} #1}}

\begin{document}
	
	\title{Experimental Realization of Criticality-Enhanced Global Quantum Sensing via Non-Equilibrium Dynamics}
	
	\date{\today}

\author{Yefei Yu}
\thanks{These authors contribute equally}
\affiliation{Beijing Academy of Quantum Information Sciences, Beijing 100193, China}
\author{Ran Liu}
\thanks{These authors contribute equally}
\affiliation{Institute of Quantum Precision Measurement, State Key Laboratory of Radio Frequency Heterogeneous Integration, College of Physics and Optoelectronic Engineering, Shenzhen University, Shenzhen 518060, China}

\author{Guangming Xue}
\affiliation{Beijing Academy of Quantum Information Sciences, Beijing 100193, China}
\affiliation{Hefei National Laboratory, Hefei 230088, China}

\author{Chuhong Yang}
\affiliation{Beijing Academy of Quantum Information Sciences, Beijing 100193, China}

\author{Chenlu Wang}
\affiliation{Beijing Academy of Quantum Information Sciences, Beijing 100193, China}

\author{Jingning Zhang}
\affiliation{Beijing Academy of Quantum Information Sciences, Beijing 100193, China}

\author{Jiangyu Cui}
\email{cjy1991@mail.ustc.edu.cn}
\affiliation{Department of Modern Physics,
University of Science and Technology of China, Hefei 230026, China}

\author{Xiaodong Yang}
\email{yangxd@szu.edu.cn}
\affiliation{Institute of Quantum Precision Measurement, State Key Laboratory of Radio Frequency Heterogeneous Integration, College of Physics and Optoelectronic Engineering, Shenzhen University, Shenzhen 518060, China}

\author{Jun Li}
\email{lijunquantum@szu.edu.cn}
\affiliation{Institute of Quantum Precision Measurement, State Key Laboratory of Radio Frequency Heterogeneous Integration, College of Physics and Optoelectronic Engineering, Shenzhen University, Shenzhen 518060, China}

\author{Jiaxiu Han}
\email{hanjx@baqis.ac.cn}
\affiliation{Beijing Academy of Quantum Information Sciences, Beijing 100193, China}
\affiliation{Hefei National Laboratory, Hefei 230088, China}
 
\author{Haifeng Yu}
\affiliation{Beijing Academy of Quantum Information Sciences, Beijing 100193, China}
\affiliation{Hefei National Laboratory, Hefei 230088, China}
	
\begin{abstract}
Quantum critical systems offer promising advancements in quantum sensing and metrology, yet face limitations like critical slowing down and a restricted criticality-enhanced region. 
Here, we introduce a critical sensing scheme that mitigate critical slowing down by leveraging the non-equilibrium dynamics of a perturbed Ising spin model, coupled with an adaptive strategy to enlarge its sensing interval. 
We validate the proposed scheme on a superconducting quantum processor and demonstrate that our scheme achieves a Heisenberg scaling with respect to the encoding duration. 
Additionally, the adaptive strategy tunes the model to operate near its critical point with limited prior information about the parameter, enabling what is known as global sensing. 
Our work showcases the metrological applications empowered by non-equilibrium critical dynamics and hence opens up a pathway for devising critical quantum sensors.  
 
\end{abstract}
\maketitle 
\textit{Introduction.---}
Quantum metrology exploits quantum features to achieve parameter estimation precision that surpasses the limits of classical methods \cite{giovannetti2004,Giovannetti2006}. 
Beyond quantum entanglement and squeezing, interest in using quantum criticality for metrology has grown in recent years. 
Quantum criticality refers to sudden changes in a quantum system's properties during a phase transition. 
At the critical point, a system's heightened sensitivity to changes in Hamiltonian parameters provides a valuable resource for quantum metrology \cite{Invernizzi2008,Zanardi2008,Rams2018,Frerot2018,Garbe2020,Mishra2021,Hotter2024,Mukhopadhyay2024,montenegro2024reviewquantummetrologysensing}. 
Moreover, critical metrology protocols offer inherent robustness to decoherence and aid in preparing probe states \cite{Raghunandan2018,Chen2021,Garbe2022,Alushi2024}. 
These features make criticality-based quantum metrology promising for developing sensors with quantum-enhanced precision.

Despite its theoretical appeal, only a few preliminary experimental attempts exist, including those in nuclear spin systems \cite{liu2021} and Rydberg atoms \cite{ding2022}. Two major obstacles limit its further practical application. 
First, most previous research relies on equilibrium states close to a quantum phase transition (QPT). 
Yet, as the system approaches the transition point, susceptibility diverges, and notably, the time required to complete the adiabatic ramp also diverges \cite{Garbe2020,liu2021,ying2022}. 
This phenomenon, known as critical slowing down, severely degrades sensitivity.
Second, the criticality induced precision enhancement occurs only within a very small parameter region \cite{Rams2018,Fisher1972,Montenegro2021}. 
It implies that near-perfect prior knowledge of the unknown parameter is required to operate the quantum sensor near the critical point. 
In other words, it is difficult to apply critical quantum metrology in the context of global sensing \cite{helstrom1976,holevo2011,Degen2017,mukhopadhyay2024currenttrendsglobalquantum}, in which the unknown parameter varies over a wide interval.

To address these limitations, we integrate non-equilibrium dynamics \cite{Macieszczak2016,Chu2021,Guan2021,Gonzalez2022,Gietka2022,Zhou2023} and adaptive parameter estimation strategies \cite{Zanardi2008,Montenegro2021,Salvia2023} into critical sensing. 
Specifically, by leveraging non-equilibrium dynamics, we are able to mitigate the critical slowing down that occurs during the preparation of equilibrium states, thereby significantly enhancing sensitivity. 
Additionally, the adaptive strategy, which incorporates feedback control into Bayesian estimation, broadens the effective sensing interval. 
Furthermore, we utilize the first-order QPT of a perturbed Ising model to demonstrate critical quantum sensing. 
Within this framework, the finite-size system, even the minimal one comprising only two interacting qubits, suffices to exhibit marked criticality-enhanced precision, which is highly amenable to the practical realization.
The experiments are conducted on a superconducting processor, whose high-quality quantum operations and readout make it an ideal platform for implementing our sensing scheme. 
The experimental results show that our scheme achieves Heisenberg-scaling precision at the critical point, and significantly surpasses the precision obtained by numerical calculations based on the equilibrium scheme.
Moreover, our online adaptive strategy allows us to asymptotically approach the quantum Cram{\'e}r-Rao bound (QCRB) \cite{helstrom1976,Braunstein1994,petz2011}, even with limited prior knowledge of the unknown parameter.

\begin{figure}[t]	 
		\begin{center}
 \includegraphics[width=0.95\linewidth]{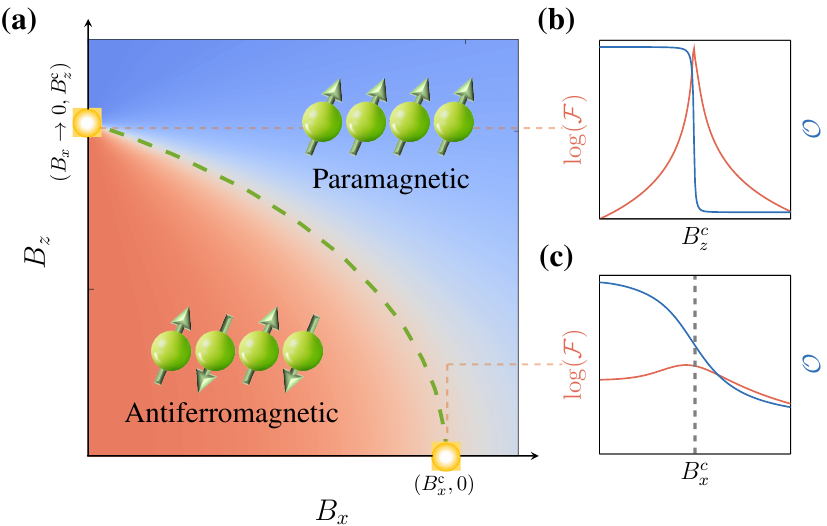} 
\caption{Schematic phase diagram and qualitative metrology potential of a finite-size one-dimensional antiferromagnetic Ising chain under control parameters $B_x$ and $B_z$. (a) Color gradients indicate the antiferromagnetic order parameter $\mathcal{O}$. The second-order phase transition is marked by a green dashed line, except at the first-order transition point $(B_x, B_z) = (0, 1)$. (b) Near the first-order transition at $(B_x, B_z) = (0, 1)$, there is a significant increase in quantum Fisher information $\mathcal{F}$, and a sharp change in $\mathcal{O}$. (c) Near the second-order transition at $(B_x, B_z) = (1/2, 0)$, $\mathcal{F}$ is relatively low, and $\mathcal{O}$ changes smoothly.}\label{Fig1}
		\end{center} 
\end{figure}

\textit{Critical quantum sensing scheme.---}We consider a finite-size one-dimensional antiferromagnetic Ising chain with periodic boundary condition, described by the Hamiltonian
\begin{equation}\label{H1}
	H=\frac{J}{2}\sum_{i=1}^{N} \sigma_z^i\sigma_z^{i+1}+ \sum_{i=1}^N \left(B_z\sigma_z^i+B_x\sigma_x^i \right),
\end{equation}
where $N$ is the number of spins, $\sigma_\mu^i$ ($\mu\in\{x,y,z\}$) are Pauli operators at the $i$th site, $J$ denotes the nearest-neighbor antiferromagnetic coupling strength (hereafter we set $J=1$ as the energy unit), $B_x$ and $B_z$ are the transverse and longitudinal field strengths, respectively. 
The competing antiferromagnetic coupling and external field cause this model to undergo a QPT between the paramagnetic and antiferromagnetic phases at zero temperature \cite{simon2011}. 
The two phases are separated by a critical line corresponding to a second-order phase transition, except at the first-order transition point where $(B_x, B_z) = (0, 1)$ \cite{Ovchinnikov2003} (refer to Fig.~\ref{Fig1}(a) for the schematic illustration).

We investigate the metrological performace in the vicinity of the critical points, employing the quantum Fisher information (QFI or $\mathcal{F}$) as a metric to assess the metrological potential. 
The QFI dictates the precision limit via the QCRB, i.e., $\Delta x\ge1/\sqrt{\nu\mathcal F}$ \cite{helstrom1976,Braunstein1994,petz2011}, where $x$ represents the parameter to be estimated (e.g., $B_x$ or $B_z$), and $\nu$ is the number of independent measurements. 
We initially focus on the second-order transition, using the exactly solvable transition point $(B_x,B_z)=(1/2,0)$ as a touchstone \cite{Sachdev2011}, whose potential applications in criticality-enhanced sensing have been intensively studied \cite{Invernizzi2008,Boyajian2016,Frerot2018,Rams2018}. 
In the context of finite-size system, the phase transition at $(B_x,B_z)=(1/2,0)$ is characterized by a gradual crossover, where the order parameter $\mathcal O$ varies smoothly, and the value of $\mathcal{F}$ remains small, as shown in Fig.~\ref{Fig1}(c). The feature leads to a diminished sensing benefit from quantum criticality.
Considering this, we utilize the first-order transition point $(B_x,B_z)=(0,1)$ in our sensing scheme. At this point, the energy gap increases linearly with the perturbation field $B_x$, allowing fine-tuning of the critical behavior \cite{Ovchinnikov2003,Rossini2018,Bonfim2019}. 
$B_x\to0$ results in a divergent $\mathcal{F}$ near the critical point, which enables the precise estimation of $B_z$ in a finite-size system, as shown in Fig.~\ref{Fig1}(b). 
One can find more details in the Supplemental Material \cite{supp}.

\begin{figure*} 	
\begin{center}
			\includegraphics[width=0.85\linewidth]{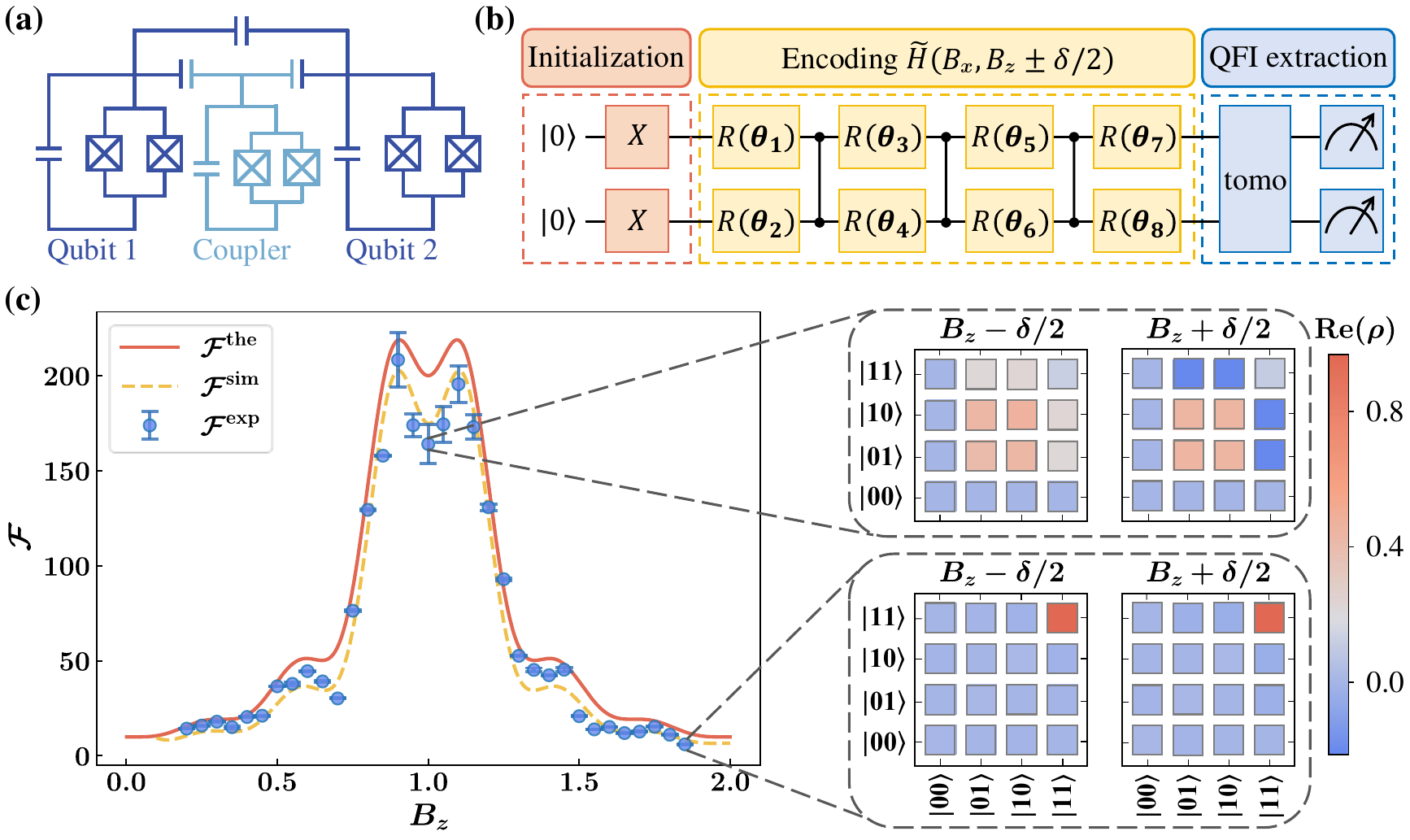} 
			\caption{Experimental demonstration of criticality-enhanced sensing. (a) Schematic circuit for the superconducting quantum sensor, comprising two transmon qubits and a tunable coupler. (b) Quantum circuit for preparing the initial state, encoding by the evolutions under $\widetilde{H}(B_x,B_z\pm \delta/2)$, and measuring the quenched state. Each single-qubit rotational gate $R(\boldsymbol{\theta}_i)$ is parameterized by three Euler angles $\boldsymbol{\theta}_i=(\alpha_i,\beta_i,\gamma_i)$. (c) $\mathcal{F}$ as a function of $B_z$. Here $\mathcal F^\text{the}$ and $\mathcal F^\text{sim}$ represent the theoretical $\mathcal{F}$ and the simulated $\mathcal{F}$ with $\delta=0.1$, respectively. The marker and error bar indicate the bootstrapped mean and standard deviation of the experimental QFI $\mathcal F^\text{exp}$. The experimentally reconstructed density matrices of the neighboring quantum states in the parameter space are shown on the right.}\label{Fig2}
		\end{center}
\end{figure*}

We leverage the non-equilibrium dynamics of the critical model for quantum sensing. 
To simplify our analysis, we focus on a two-spin model, as described by the Hamiltonian
\begin{equation}\label{H2}
	\widetilde H=B_z(\sigma_z^1+\sigma_z^2)+B_x(\sigma_x^1+\sigma_x^2)+\sigma_z^1\sigma_z^2.
\end{equation}
Initially, we prepare the system in the state $|\psi_0\rangle=|11\rangle$. 
We then implement a quantum quench to encode the parameter $B_z$, which is to be estimated, by evolving the system under $\widetilde{H}$. 
Finally we can assess the metrological potential of the quenched state $|\psi_\text{f}\rangle$ via QFI.

To facilitate the derivation of QFI, we use the exchange symmetry of $\widetilde{H}$ and select a new set of basis states, $|a\rangle=|11\rangle$, $|b\rangle=(|01\rangle+|10\rangle)/\sqrt{2}$, $|c\rangle=|00\rangle$, and $|d\rangle=(|01\rangle-|10\rangle)/\sqrt{2}$. 
Under the condition of $B_x\to0$, $\widetilde{H}$ can be approximated by an effective Hamiltonian within a subspace spanned by $\{|a\rangle,|b\rangle\}$ \cite{Peng2005,Zhang2008}. 
Consequently, we can approximate $|\psi_\text{f}\rangle$ as a superposition of the states $|a\rangle$ and $|b\rangle$, which in turn allows us to explicitly derive an approximate expression for the QFI.
The dominant term of QFI near the critical point is \cite{supp}
\begin{equation}\label{domFq}
	\mathcal F(t)\sim\frac{16B_x^4}{\Omega^6}\sin^4\Omega t,
\end{equation}
where $\Omega=\sqrt{\left(1-B_z\right)^2+2B_x^2}$, and $t$ is the encoding duration. 
At the critical point $B_z=1$, $\mathcal F(t)$ attains its first maximum at $T=\pi/\Delta_c$, with $\Delta_c=2\sqrt{2}B_x$ representing the corresponding energy gap. 
Consequently, the encoding duration is optimally set to $T$.
Diminishing $B_x$ leads to a proportional decrease in $\Delta_c$, which results in an enhanced QFI for the estimation of $B_z$.
The QFI at the critical point can also be expressed as a function of encoding duration $T$, as given by 
\begin{equation}\label{HL}
  	\mathcal F|_{B_z=1}(T)=\frac{16}{\pi^2}T^2,
\end{equation}
which illustrates a Heisenberg scaling characterized by a prefactor $\chi=16/\pi^2$. 
It is worth mentioning that, with a carefully designed adiabatic path, the Heisenberg scaling has also been achieved via preparing the ground state at the critical point \cite{Rams2018,Garbe2020,liu2021}. 
Nevertheless, the requirement for a slow adiabatic evolution, as stipulated by the adiabatic theorem, results in $\chi\ll1$, thereby limiting the practical metrology.

\textit{Experimental setup and procedure.---}We demonstrate the protocol utilizing two superconducting transmon qubits, as depicted in Fig.~\ref{Fig2}(a). 
A tunable coupler facilitates high-contrast qubit interaction switching, which is crucial for achieving high-fidelity single- and two-qubit gates. 
The average fidelity of the single-qubit gates for each of the qubits is 99.92$\%$, and the fidelity of the controlled-Z (CZ) gate is 99.1$\%$. 
The qubits have readout fidelities of 95.3$\%$ and 96.1$\%$, and we apply extra readout corrections to boost measurement precision.
Further details are provided in \cite{supp}.

To obtain the quenched state $|\psi_\text{f}\rangle$, we initialize the system to $|\psi_0\rangle$ using X gates on both qubits and let it evolve under $\widetilde{H}$ for a duration $T$, as shown in Fig.~\ref{Fig2}(b).
The evolution can generally be approximated through Trotter-Suzuki decomposition \cite{suzuki1993}, yet given that our demonstration involves only two qubits, we opt for a more precise realization by compiling the desired transformation into a simplified quantum circuit, which includes 3 CZ gates and 8 single-qubit gates.
Adjusting $B_x$ and $B_z$ values only affects the rotation angles $\boldsymbol{\theta}$ of the single-qubit gates, without altering the circuit's structure.
We then use $|\psi_\text{f}\rangle$ to assess metrological performance with various measurement strategies.

\textit{Enhanced Precision Near the Critical Point.---} To evaluate the metrological potential near the critical point, we set $B_x$ to 0.1 and vary $B_z$ from 0 to 2, anticipating an increase in QFI around $B_z^\text{c} = 1$. 
The QFI for the quenched state $\rho(B_z)$ is calculated by
\begin{equation}\label{Fqexp}
	\mathcal{F}=4\lim_{\delta\to0}\frac{D^2_{\mathrm{B}}\left[\rho(B_{z}-\delta/2), \rho(B_{z}+\delta/2)\right]}{\delta^2},
\end{equation}
where $D_{\mathrm{B}}\left[\rho_1, \rho_2\right]=\sqrt{2-2 \text{Tr}\sqrt{\sqrt{\rho_1}\rho_2\sqrt{\rho_1}}}$ is the Bures distance \cite{helstrom1969,Braunstein1994} between $\rho_1$ and $\rho_2$. 
Equation \eqref{Fqexp} indicates that a smaller $\delta$ provides a more precise approximation of $\mathcal{F}$, but an extremely small $\delta$ could amplify experimental errors in $D_{\mathrm{B}}$. 
Thus we pragmatically select $\delta = 0.1$ to balance the experimental feasibility. 
Based on this, we prepare the quenched states under $\widetilde{H}(0.1,B_z\pm\delta/2)$ and reconstruct $\rho(B_z\pm\delta/2)$ via quantum state tomography for $\mathcal{F}$ computation, as shown in the quantum circuit in Fig.~\ref{Fig2}(b).

The experimental results depicted in Fig.~\ref{Fig2}(c) show a significant increase in $\mathcal{F}$ near the critical point, which is characterized by a distinctive double-peaked pattern attributed to the non-equilibrium method \cite{Hotter2024}, confirming enhanced precision for $B_z$ estimation.
The experimental data $\mathcal{F}^\text{exp}$ closely align with numerical simulations $\mathcal{F}^\text{sim}$ for $\delta = 0.1$. 
We employ Pauli twirling \cite{Bennett1996,Geller2013} within the circuit to prevent the buildup of coherent errors and average out the noise impact.
The median state fidelity of the quenched states, measured at $97.35\%$, closely matches the fidelity inferred from gate fidelities.
A minor discrepancy between $\mathcal{F}^\text{sim}$ and theoretical predictions $\mathcal{F}^\text{the}$ is attributed to the non-zero $\delta$.
In addition, the criticality-enhanced precision is also corroborated through quantum state distinguishability \cite{Fernandez2017, gefen2019}, as illustrated by the density matrices in the right panel of Fig.~\ref{Fig2}(c). 
The states near the critical point, $\rho\left(1\pm\delta/2\right)$, exhibit a high degree of distinguishability, with $D_{\mathrm{B}}=0.64$. 
In contrast, the states $\rho\left(1.85\pm\delta/2\right)$ show only a minor difference, with $D_{\mathrm{B}}=0.12$.

\begin{figure}	
		\begin{center}
			\includegraphics[scale=0.36]{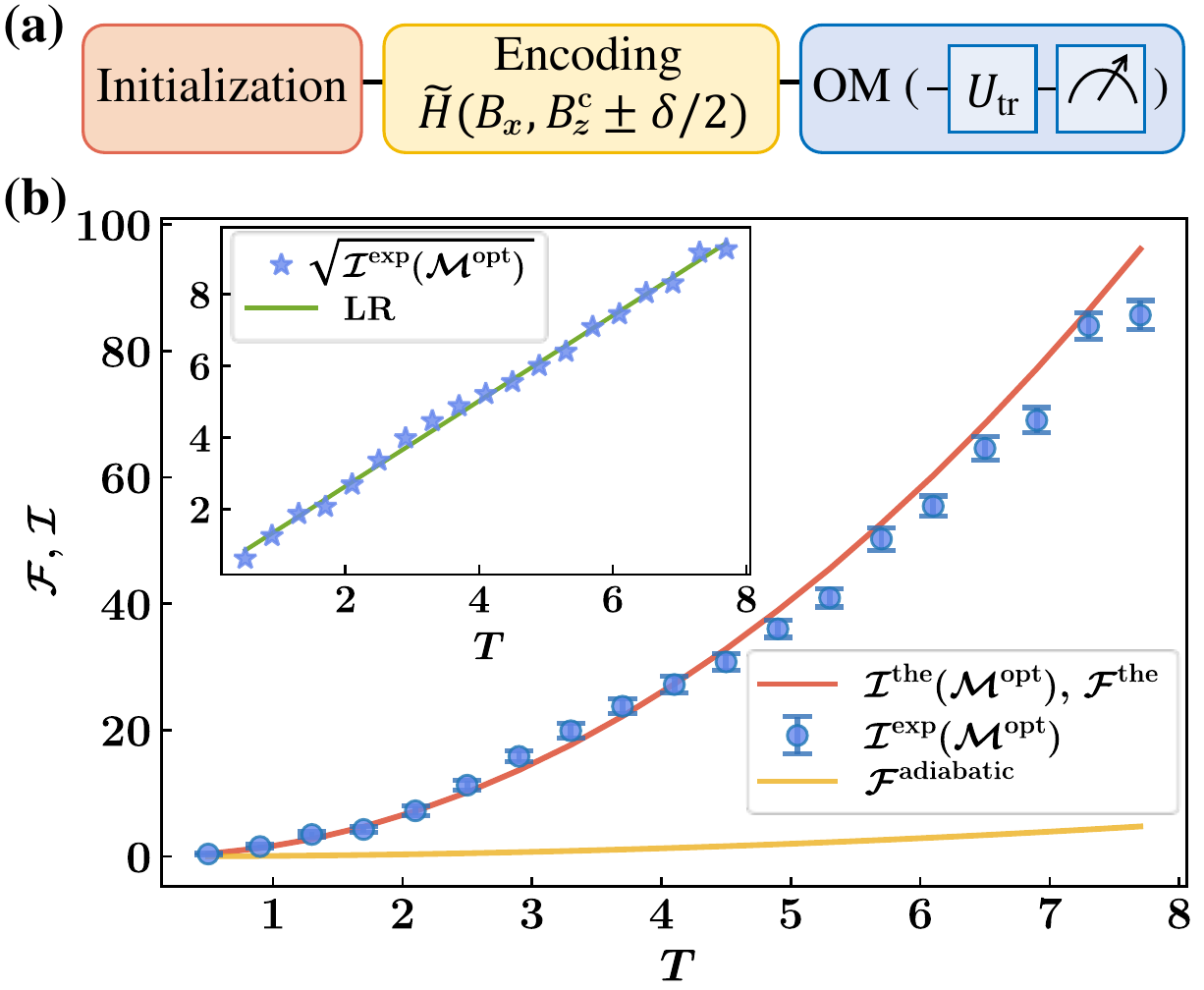} 
			\caption{Scaling behavior at the critical point.
	 (a) Experimental protocol. OM refers to the optimal measurement under $\mathcal{M}^\text{opt}$, which includes a unitary transformation $U_\text{tr}$ to convert the readout basis and a projective measurement. (b) CFI or QFI as a function of  $T$. $\mathcal F^\text{the}$ represents the theoretical QFI, which equals the theoretical CFI under optimal measurements, i.e., $\mathcal I^\text{the}(\mathcal M^\text{opt})$. $\mathcal I^\text{exp}(\mathcal M^\text{opt})$ is the experimentally extracted CFI. $\mathcal{F}^{\text{adiabatic}}$ shows simulations from the conventional critical sensing scheme based on preparing ground states. The inset shows the linear regression (LR) on $\sqrt{\mathcal I^\text{exp}(\mathcal M^\text{opt})}$, with a coefficient of determination $R^2=99.6\%$.}\label{Fig3}
		\end{center}
\end{figure}

\textit{The scaling behavior at the critical point.---}The QFI delineates the metrological potential, establishing the ultimate precision limit in quantum metrology. For a specific measurement strategy, its sensing performance is typically assessed using the classical Fisher information (CFI). Upon employing the optimal measurement basis \(\mathcal{M}^{\text{opt}}\), the CFI reaches its peak value, which coincides with the QFI \cite{helstrom1969,Braunstein1994,holevo2011}.
Based on the symmetric logarithmic derivative \cite{Braunstein1994}, the optimal measurement directions at the critical point are given by $\mathcal M^{\text{opt}}=\{|v^\text{opt}_m\rangle\langle v^\text{opt}_m|\}_{m=1}^4$ with $|v^\text{opt}_{1,2}\rangle=(|a\rangle\pm|b\rangle)/\sqrt2$, $|v^\text{opt}_{3,4}\rangle=(|c\rangle\pm|d\rangle)/\sqrt2$.
To implement the optimal measurement on the qubits, a unitary transformation $U_\text{tr}=\sum_m|v_m^\text{exp}\rangle\langle v^\text{opt}_m|$ should be applied before the projective measurement to adjust the readout basis, where ${|v_m^\text{exp}\rangle}={|00\rangle,|01\rangle,|10\rangle,|11\rangle}$.
The CFI is then derived from 
\begin{equation}
	\label{CFI}
	\mathcal I(\mathcal M^{\text{opt}})=\sum_m[\partial_{B_z}p(m|B_z)]^2/p(m|B_z),
\end{equation}
where $p(m|B_z)=\text{Tr}[\rho_\text{f}(B_z)|v^\text{opt}_m\rangle\langle v^\text{opt}_m|]$ represents the conditional probability of outcome $m$.

We experimentally investigate the relationship between precision and encoding duration $T$. 
With $B_z$ fixed at the critical point, we systematically adjust $B_x$ from 2.22 to 0.14, which corresponds to an increase in $T$ from 0.50 to 7.93.
The derivative $\partial_{B_z}p(m|B_z)$ is approximated using the
finite difference method: $\partial_{B_z}p(m|B_z)\approx[p(m|B_z+\delta/2)-p(m|B_z-\delta/2)]/\delta$, with $\delta$ set to 0.06 to balance experimental and approximation errors. 
Following the procedure outlined in Fig.~\ref{Fig3}(a), we conduct the experiments and statistically determine the conditional probabilities from $5.12 \times 10^5$ experimental shots.
Experimental CFI, $\mathcal I^\text{exp}(\mathcal M^\text{opt})$, matches theoretical values, $\mathcal I^\text{the}(\mathcal M^\text{opt})$ or $\mathcal F$, as shown in Fig.~\ref{Fig3}(b). 
Linear regression on $\sqrt{\mathcal I^\text{exp}(\mathcal M^\text{opt})}$ gives an $R$-squared of 99.6\%, indicating a quadratic increase in $\mathcal I^\text{exp}(\mathcal M^\text{opt})$ with $T$, which demonstrates the Heisenberg scaling of criticality-enhanced precision.
For comparison, we also present the QFI simulated from the adiabatic passage in Fig.~\ref{Fig3}(b) \cite{supp}, which exhibits significantly lower precision than our results.

\emph{Global sensing with adaptive strategy.---}As demonstrated, precision is heightened near the critical point. 
In local sensing, the initial estimate of $B_z$, denoted as $\hat{B}_z$, is very close to the true value. 
We can apply a control field $B_z^\text{ctrl} = B_z^\text{c} - \hat{B}_z$ to steer the system towards the critical point, enhancing the precision of $B_z$ measurement. 
However, in global sensing, with $B_z$ spanning a wide range and scarce prior knowledge, leveraging criticality is much more challenging.

\begin{figure}	
		\begin{center}
			\includegraphics[scale=0.43]{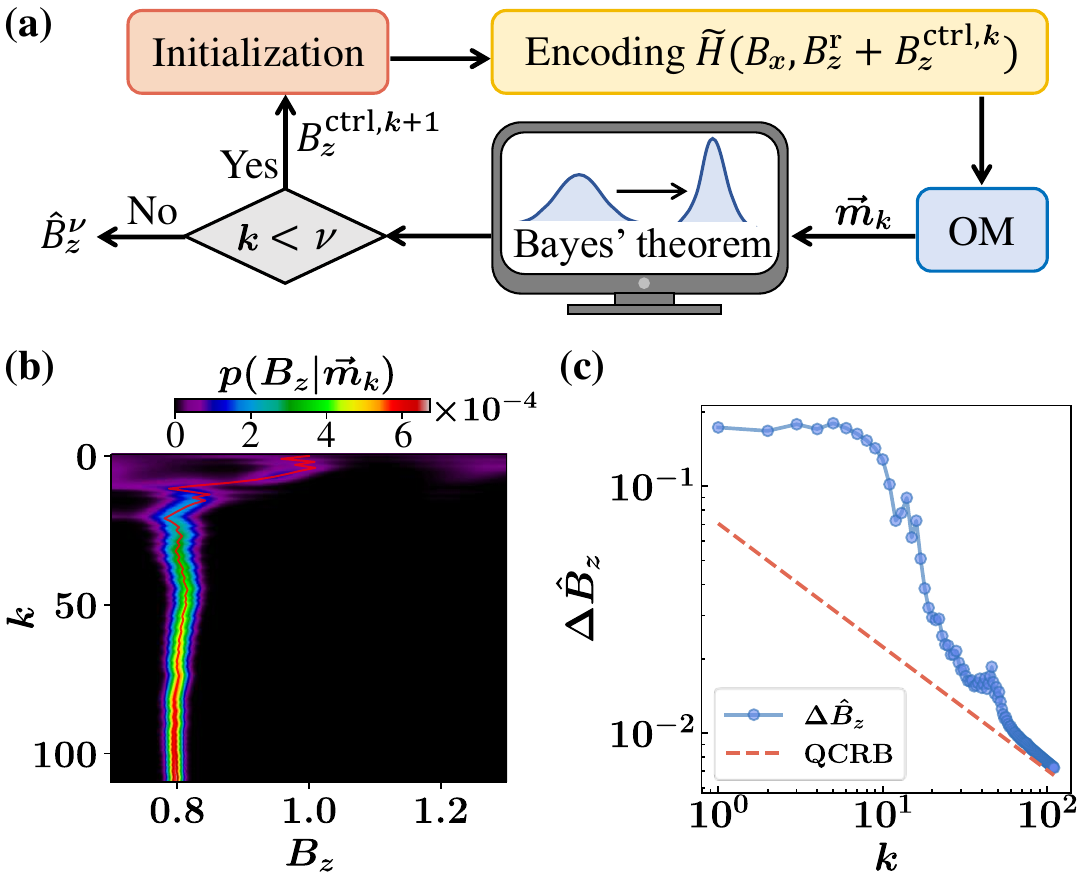} 
			\caption{Experimental demonstration of the adaptive strategy for global sensing. (a) Flowchart of the adaptive strategy. (b) The trajectory of the posterior probability distribution $p(B_z|\vec{m}_k)$ in a single adaptive estimation of $B_z$. A total of 110 measurements, denoted by $\nu$, are conducted. The red line represents the estimated $B_z$ during the adaptive estimation. (c) Measured uncertainty of the parameter. The blue dots represent the standard deviation of experimental estimation $\hat B_z^k$, which asymptotically approaches the QCRB at the critical point, indicated by the red dashed line.}\label{Fig4}
		\end{center}
\end{figure}

In our study, we employ an adaptive strategy \cite{Zanardi2008, Montenegro2021,Salvia2023} that integrates feedback control with Bayesian estimation to guide the quantum system to its critical region for sensing \cite{supp}. 
We assume the real $B_z$ is $B_z^{\text{r}}=0.8$, not near the critical point, and our prior knowledge places $B_z$ in [0.7, 1.3]. 
Starting with a flat prior distribution $p_0(B_z)=1/(B_z^\text{max}-B_z^\text{min})$, the initial $B_z$ estimation is $\hat B_z^0=\int B_z p_0(B_z)dB_z$. 
The experimental flowchart is shown in Fig.~\ref{Fig4}(a), which involves $\nu$ single-shot measurements. 
For each $k$-th measurement, we initialize the system and quench under $\widetilde{H}(B_x=0.1, B_z^{\text{r}} + B_z^{\text{ctrl,}k})$, with $B_z^{\text{ctrl,}k} = B_z^\text{c} - \hat{B}_z^{k-1}$ adjusting the system towards the critical point, based on the $(k-1)$th estimated value $\hat{B}_z^{k-1}$. 
We measure the quenched state in the optimal directions for the critical point. 
Given the experimental outcomes of the first $k$ measurement $\vec{m}_k=\{m_1,m_2,...m_k\}_{m=1}^4$, we update the posterior probability distribution using Bayes' theorem \cite{sivia2006}
\begin{equation}
	p(B_z|\vec{m}_{k})=\mathcal Np(m_k|B_z+B_z^{\text{ctrl,}k})p(B_z|\vec{m}_{k-1}),
\end{equation}  
where $p(m_k|B_z+B_z^{\text{ctrl,}k})=\text{Tr}[\rho(B_z+B_z^{\text{ctrl,}k})|v^\text{opt}_m\rangle\langle v^\text{opt}_m|]$ represents the conditional probability of observing the outcome $m_k$, and $\mathcal{N}$ is the normalization constant. 
Then we estimate $B_z$ through $\hat{B}_z^k=\int B_z p(B_z|\vec{m}_{k})dB_z$ and determine the control field $B_z^{\text{ctrl,}k+1}$ for the next measurement.

The experimental trajectory of $p(B_z|\vec{m}_{k})$ is shown in Fig.~\ref{Fig4}(b). 
As the number of adaptive measurements increases, the estimated $B_z$ converges to the true value $B_z^{\text{r}}$.
Meanwhile, as the total longitudinal field $B_z^{\text{r}}+B_z^{\text{ctrl},k}$ approaches the critical point, it yields a progressive enhancement in measurement sensitivity. This trend is further corroborated by the rapidly decreasing uncertainty of $p(B_z|\vec{m}_{k})$.
The precision of the experimental estimation can be quantified by the standard deviation of the posterior distribution: $\Delta\hat B_z^k=\left[\int p(B_z|\vec{m}_{k})(B_z-\hat B_z^k)^2dB_z\right]^{1/2}$ \cite{li2018}. 
As shown in Fig.~\ref{Fig4}(c), $\Delta\hat B_z^k$ decreases dramatically with the increasing number of adaptive measurements, asymptotically approaching the fundamental precision bound---QCRB at the critical point $B_z^\text{c}$.

\textit{Summary---}We experimentally validate the criticality-enhanced global quantum sensing by employing non-equilibrium dynamics and adaptive estimation strategies on a superconducting quantum processor. 
To the best of our knowledge, this study represents the first experimental achievement of Heisenberg scaling in the field of critical quantum sensing via non-equilibrium dynamics. 
Additionally, it is the first time that critical quantum sensing has reached the QCRB in global sensing.
Our work offers a promising approach to overcoming the key challenges in quantum metrology, and opens up a new avenue for the development of practical critical quantum sensors.

\begin{acknowledgments}
This work is supported by National Natural Science Foundation of China (Grants No. 92365206,  No. 12204230, No. 1212200199, No. 92065111, No. 12441502, No. 11975117, No. 12404554), Innovation Program for Quantum Science and Technology (No. 2021ZD0301802), Shenzhen Science and Technology Program (Grant No. RCYX20200714114522109), Guangdong
Basic and Applied Basic Research Foundation (Grant
No. 2021B1515020070), China Postdoctoral Science Foundation (2024M762114), Postdoctoral Fellowship Program of CPSF (Grant No. GZC20231727).
\end{acknowledgments}

	\bibliographystyle{apsrev4-2}
	\bibliography{Reference}

\end{document}

\title{Experimental Realization of Criticality-Enhanced Global Quantum Sensing via Non-Equilibrium Dynamics: Supplemental Material}

	\date{\today}
\author{Yefei Yu}
\thanks{These authors contribute equally}
\affiliation{Beijing Academy of Quantum Information Sciences, Beijing 100193, China}
\author{Ran Liu}
\thanks{These authors contribute equally}
\affiliation{Institute of Quantum Precision Measurement, State Key Laboratory of Radio Frequency Heterogeneous Integration, Shenzhen University, Shenzhen 518060, China}

\author{Guangming Xue}
\affiliation{Beijing Academy of Quantum Information Sciences, Beijing 100193, China}
\affiliation{Hefei National Laboratory, Hefei 230088, China}

\author{Chuhong Yang}
\affiliation{Beijing Academy of Quantum Information Sciences, Beijing 100193, China}

\author{Chenlu Wang}
\affiliation{Beijing Academy of Quantum Information Sciences, Beijing 100193, China}

\author{Jingning Zhang}
\affiliation{Beijing Academy of Quantum Information Sciences, Beijing 100193, China}

\author{Jiangyu Cui}
\email{cjy1991@mail.ustc.edu.cn}
\affiliation{Institute of Quantum Precision Measurement, State Key Laboratory of Radio Frequency Heterogeneous Integration, Shenzhen University, Shenzhen 518060, China}

\affiliation{College of Physics and Optoelectronic Engineering, Shenzhen University, Shenzhen 518060, China}

\author{Xiaodong Yang}
\email{yangxd@szu.edu.cn}
\affiliation{Institute of Quantum Precision Measurement, State Key Laboratory of Radio Frequency Heterogeneous Integration, Shenzhen University, Shenzhen 518060, China}

\affiliation{College of Physics and Optoelectronic Engineering, Shenzhen University, Shenzhen 518060, China} 

\author{Jun Li}
\email{lijunquantum@szu.edu.cn}
\affiliation{Institute of Quantum Precision Measurement, State Key Laboratory of Radio Frequency Heterogeneous Integration, Shenzhen University, Shenzhen 518060, China}

\affiliation{College of Physics and Optoelectronic Engineering, Shenzhen University, Shenzhen 518060, China}

\author{Jiaxiu Han}
\email{hanjx@baqis.ac.cn}
\affiliation{Beijing Academy of Quantum Information Sciences, Beijing 100193, China}
\affiliation{Hefei National Laboratory, Hefei 230088, China}
 
\author{Haifeng Yu}
\affiliation{Beijing Academy of Quantum Information Sciences, Beijing 100193, China}
\affiliation{Hefei National Laboratory, Hefei 230088, China}
	\maketitle
	
\maketitle

\tableofcontents
\section{Quantum Phase Transitions in 1D Antiferromagnetic Ising Chains}	
The Hamiltonian of the one-dimensional antiferromagnetic Ising chain is
\begin{equation}\label{Ham}
	H=\frac{J}{2} \sum_{i=1}^N \sigma_z^i \sigma_z^{i+1}+B_z \sigma_z^i+B_x \sigma_x^i.
\end{equation}
where $J>0$ characterizes the strength of the nearest-neighbor antiferromagnetic coupling, $B_x$ and $B_z$ are the strengths of the transverse and longitudinal magnetic fields, respectively, and a periodic boundary condition is imposed. In the following, we set $J=1$ as the energy unit. The properties of QPT can be characterized via the change of antiferromagnetic order parameter defined as \cite{Sachdev2002}
\begin{equation}
	\mathcal O=\left\langle\left(\frac{1}{N} \sum_i(-1)^i \frac{\sigma_z^i}{2}\right)^2\right\rangle
\end{equation}
with $\langle\cdot\rangle$ denoting the expectation value for the ground state $|g\rangle$. Moreover, quantum Fisher information (QFI), which is defined as
 \begin{equation}\label{QFI}
 	\mathcal F(|g\rangle)=4(\langle\partial_{x}g|\partial_{x}g\rangle-|\langle g|\partial_{x}g\rangle|^2), 
 \end{equation}
 is closely related to the susceptibility of quantum state to the change of system parameter $x$ ($x$ is $B_x$ or $B_z$ in this case) and can also be employed to study the quantum critical phenomenon \cite{Rams2018}.

Since the Hamiltonian in Eq. \eqref{Ham} is not exactly solvable, its critical line has been obtained using the density-matrix renormalization-group (DMRG) technique \cite{Ovchinnikov2003}.  Here we specifically study the properties of QPT in the vincinity of two special quantum critical points, i.e., $(B_x=0.5,B_z=0)$ and $(B_x=0,B_z=1)$, by observing antiferromagnetic order parameter $\mathcal O$ and QFI $\mathcal F$.
\begin{figure*}
		\begin{center}
			\includegraphics[scale=1.6]{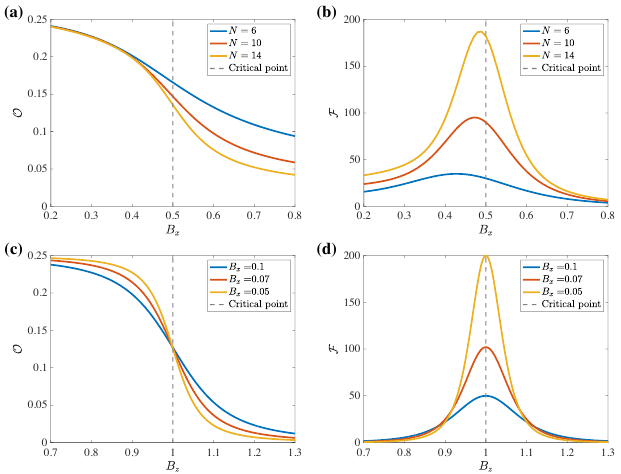} 
			\caption{(a) Order parameter $\mathcal O$ and (b) QFI $\mathcal F$ in the second-order phase transition, where $B_x=0.5$ is the critical point. Due to the finite-size cutoff, the sharp phase transition at critical point is rendered to a smooth crossover. (c) Order parameter $\mathcal O$ and (d) QFI $\mathcal F$ in the first-order phase transition, where $B_z=1$ is the multicritical point. The behavior of phase transition can be tuned via the scaling of system parameter $B_z$, thus exhibiting a sharp transition in finite-size system.}\label{FigS1}
		\end{center}
\end{figure*}
\begin{itemize}
	\item \textbf{Critical point of the second-order phase transition $(B_x=0.5,B_z=0)$}

	When $B_z=0$, Eq. \eqref{Ham} becomes a transverse field Ising model
	\begin{equation}\label{TFIM}
		H=\frac{J}{2} \sum_i \sigma_z^i \sigma_z^{i+1}+B_x \sigma_x^i.
	\end{equation}
	At the critical point $(B_x=0.5,B_z=0)$, a continuous second-order phase transition occurs in the thermodynamic limit i.e., $N\to\infty$. To facilitate analysis, we can diagonalize this model with a Jordan-Wigner tranformation followed by a Bogliubov transformation \cite{Sachdev2011}.
	
	Firstly, we express Eq. \eqref{TFIM} with fermionic creation and annihilation operators with the maping
	\begin{eqnarray}
		\sigma_i^x&=&1-2 c_i^{\dagger} c_i \\
\sigma_i^z&=&-\prod_{j<i}\left(1-2 c_j^{\dagger} c_j\right)\left(c_j+c_j^{\dagger}\right),
	\end{eqnarray}
where $c_i$ and $c_i^\dagger$ are annihilation and creation operators for the spinless fermions on site $i$ and satisfy the anti-commutation relations. We then performs the Fourier transformation
	$c_q=(1/\sqrt{N}) \sum_q e^{-iqj} c_j$. The Hamiltonian is then decomposed to a matrix form as
\begin{equation}\label{ftH}
	\tilde{H}=\frac{1}{2}\sum_q\left(\begin{array}{ll}
c_q^{\dagger} & c_{-q}
\end{array}\right)\left(\begin{array}{cc}
2B_x-\cos q & i \sin q \\
-i \sin q & -(2B_x-\cos q)
\end{array}\right)\binom{c_q}{c_{-q}^{\dagger}}.
\end{equation}
We finally diagonalize Eq. \eqref{ftH} by performing a Bogoliubov transformation on the two-dimensional block of the operators of fermions $c_q,c_{-q}$ with $f_q=u_qc_q-i v_qc_{-q}^\dagger$, where $u_q$ and $v_q$ are real numbers satisfying $u_q^2+v_q^2=1$, $u_{-q}=u_q$, and $v_{-q}=-v_q$. The Hamiltonian becomes
\begin{equation}\label{H2}
	\tilde{H}=\sum_q\epsilon_q(f_q^\dagger f_q-\frac{1}{2}),
\end{equation}
where $\epsilon_q=\sqrt{1+4B_x^2-4B_x\cos q}$. The ground state of Eq. \eqref{H2} is
\begin{equation}
	|g\rangle=\prod_q \left(\cos \frac{\theta_q}{2}-i \sin \frac{\theta_q}{2} c_{-q}^{\dagger} c_q^{\dagger}\right)|\emptyset\rangle
\end{equation}
with $\tan \theta_q=\sin q/(2B_x+\cos q)$ and $|\emptyset\rangle$ being the vaccum state of fermion.
The quantum Fisher information (QFI) of $|g\rangle$ can thus be obtained according to Eq. \eqref{QFI} as
\begin{eqnarray}\label{FqIs}
	\mathcal F(|g\rangle)=\sum_q\left(\frac{\sin q}{1+4B_x^2-4B_x\cos q}\right)^2.
\end{eqnarray}

Considering the application in practical quantum system with finite size, we simulate the order parameter $\mathcal O$ and QFI $\mathcal F$ of the ground states acrossing the critical point for $N=6,10,14$. From the behavior of $\mathcal O$ in the proximity of the critical point in Fig. \ref{FigS1}(a), we can see that, due to the finite-size cutoff, the sharp phase transition at critical point is rendered to a smooth crossover. This phenomenon can also be reflected from the behavior of $\mathcal F$ in Eq. \eqref{FqIs} and Fig. \ref{FigS1}(b), in which the decay of $\mathcal F$ in a small-sized system indicates a decreasing susceptibility.

	\item \textbf{Multicritical point of the first-order phase transition ($B_x=0,B_z=1$)}
	
	When $B_x=0$, Eq. \eqref{Ham} becomes
	\begin{equation}\label{Ham1st}
	H=\frac{J}{2} \sum_{i=1}^N \sigma_z^i \sigma_z^{i+1}+B_z \sigma_z^i.
\end{equation}
This model is also exactly solvable, in which a first-order phase transition occurs at $B_z=1$. There is an abrupt transition at this critical point in finite-size systems, while the ground state is macroscopic degenerate and does not provide a precise information of $B_z$ from the perspective of quantum metrology. Here we introduce a perturbation transverse field $B_x\ll1$ to lift the degeneracy as well as encode the information of $B_z$ into the ground state. 

To analytically study the properties of QPT in this model, we focus on the minimal system consisting of two spins, i.e., 
\begin{equation}\label{HN2}
	H=B_z\left(\sigma_z^1+\sigma_z^2\right)+B_x\left(\sigma_x^1+\sigma_x^2\right)+\sigma_z^1 \sigma_z^2.
\end{equation}
Working with the subspace spanned by the two lowest energy states $\{|11\rangle,(|10\rangle+|01\rangle)/\sqrt2\}$ of Eq.~\eqref{HN2}, we give the effective Hamiltonian \cite{Zhang2008}
\begin{equation}\label{effH}
{H}_{\text{eff}}=-|B_z|\mathds{1}+(1-|B_z|)\sigma_z+\sqrt2B_x\sigma_x. 
\end{equation}
The ground state can thus be expressed as 
\begin{equation}
	|g\rangle=-\sin \frac{\theta}{2}|11\rangle+\frac{1}{\sqrt2}\cos \frac{\theta}{2}(|01\rangle+|10\rangle),
\end{equation}
where $\tan\theta=\sqrt{2}B_x/(1-B_z)$. According to Eq. \eqref{QFI}, we have the QFI
\begin{equation}\label{Fq1st}
	\mathcal F(|g\rangle)=\frac{2 B_x^2}{\left[\left(1-B_z\right)^2+2 B_x^2\right]^2}.
\end{equation}

The order parameter $\mathcal O$ and QFI $\mathcal F$ of the ground states under different $B_x=0.1,0.07$ and 0.05 are shown in Fig. \ref{FigS1}(c) and (d), respectively. From the behavior of $\mathcal O$ in the proximity of the critical point in Fig. \ref{FigS1}(c), we can see that an abrupt phase transition can be observed in a finite-size system in this regime, and it can be appropriately controlled by tuning $B_x$. This is also indicated by QFI in Eq. \eqref{Fq1st} and Fig. \ref{FigS1}(d).
\end{itemize}

From the observation of order parameter $\mathcal O$ and QFI $\mathcal F$, we find that the QPT at two typical quantum critical points in the one-dimensional antiferromagnetic Ising chain exhibits distinct properties. Due to the sharp transition in finite-size system and tunable susceptibility via the scaling of system parameter $B_x$, the multicritical point has more potential for implementing critical quantum metrology in practical quantum systems.
	
\section{QFI under the nonequilibrium quantum quench dynamics}
In our protocol, the parameter to be estimated is encoded into the initial state $|\psi_0\rangle$ via a nonequilibrium quench dynamics, i.e., $U_q=e^{-i H t}$ with $H$ being the Hamiltonian of a critical quantum system. In this case, QFI of $|\psi_\text{f}\rangle=U_q|\psi_0\rangle$ in Eq. \eqref{QFI} becomes
\begin{equation}\label{QFIquench}
	\mathcal F(|\psi_\text{f}\rangle)=4(\langle\psi_0|\partial_{B_z}U_q^\dagger\partial_{B_z}U_q|\psi_0\rangle-|\langle\psi_0|U_q^\dagger\partial_{B_z}U_q|\psi_0\rangle|^2).
\end{equation}
For the critical quantum model in Eq. \eqref{HN2}, we can work with the effective Hamiltonian in Eq. \eqref{effH}. The evolution under the identity operator only leads to a global phase, which doesn't change QFI. So in the following, we only consider the terms of $\sigma_z,\sigma_x$ in Eq. \eqref{effH} and have the evolution under the quantum quench
\begin{equation}
	U_q=e^{-i H_\text{eff}t}=r_I\mathds 1+i(\vec r\cdot\vec\sigma)
\end{equation}
where $r_I=\cos\Omega t,\vec r=(r_x,r_y,r_z)$ with
\begin{eqnarray}
	&&r_x=-\frac{h_x}{\Omega}\sin\Omega t\\\notag
&&r_y=0\\\notag
&&r_z=-\frac{h_z}{\Omega}\sin\Omega t.\notag
\end{eqnarray}
and $h_x=\sqrt2 B_x,h_z=1-B_z,\Omega=\sqrt{h_x^2+h_z^2}$. 

In our protocol, we are interested in the precise estimation of $B_z$. So we need to calculate the derivative for $B_z$, i.e.
\begin{equation}
	\partial_{B_z}U_q=v_I\mathds 1+i\vec v\cdot\vec{\sigma},
\end{equation}
where $\vec{v}=(v_x,v_y,v_z)$ with
\begin{eqnarray}
	&&v_I=-\frac{h_zt}{\Omega}\sin \Omega t\\ \notag
&&v_x=\frac{h_xh_z}{\Omega^3}\sin\Omega t-\frac{h_xh_z}{\Omega^2}t\cos \Omega t\\\notag
&&v_y=0\\\notag
&&v_z=-\frac{h_x^2}{\Omega^3}\sin \Omega t-\frac{h_z^2}{\Omega^2}t\cos \Omega t.\notag
\end{eqnarray}
For the first part $\langle\psi_0|\partial_{B_z}U_q^\dagger\partial_{B_z}U_q|\psi_0\rangle$ in Eq. \eqref{QFIquench}, we have
\begin{eqnarray}
    &&v_I^2+v_x^2+v_z^2\\
=&&\frac{h_z^2}{\Omega^2}t^2\sin^2\Omega t+\frac{h_x^2h_z^2}{\Omega^6}\sin^2\Omega t-\frac{2h_x^2h_z^2t}{\Omega^5}\sin \Omega t\cos \Omega t+\frac{h_x^2h_z^2t^2}{\Omega^4}\cos^2\Omega t\\
&&+\frac{h_x^4}{\Omega^6}\sin^2 \Omega t+\frac{2h_x^2h_z^2t}{\Omega^5}\sin \Omega t\cos \Omega t+\frac{h_z^4t^2}{\Omega^4}\cos^2 \Omega t\\
=&&\frac{h_z^2t^2}{h_z^2+h_x^2}+\frac{h_x^2}{(h_z^2+h_x^2)^2}\sin^2\Omega t.
\end{eqnarray}
For the second part $|\langle\psi_0|U_q^\dagger\partial_{B_z}U_q|\psi_0\rangle|^2$, we have
\begin{eqnarray}
    |\langle\psi_0|U_q^\dagger\partial_{B_z}U_q|\psi_0\rangle|^2&=&\langle\psi_0|U_q^\dagger\partial_{B_z}U_q|\psi_0\rangle\langle\psi_0|\partial_{B_z}U_q^\dagger U_q|\psi_0\rangle\\\notag
&=&s_I^2+\frac{1}{2}\left[(s_x\langle\sigma_x\rangle+s_y\langle\sigma_y\rangle)^2+(s_y\langle\sigma_y\rangle+s_z\langle\sigma_z\rangle)^2+(s_x\langle\sigma_x\rangle+s_z\langle\sigma_z\rangle)^2\right],\\\notag
\end{eqnarray}
where $\langle\sigma_\mu\rangle=\langle\psi_0|\sigma_\mu|\psi_0\rangle,s_I=r_Iv_I+r_xv_x+r_zv_z,\vec{s}=(s_x,s_y,s_z)$ with
\begin{eqnarray}
	s_x&=&r_Iv_x-r_xv_I,\\\notag
	s_y&=&r_zv_x-r_xv_z\\\notag
	s_z&=&r_Iv_z-r_zv_I.
\end{eqnarray}

For the easy implementation in our experiment, we consider the separable initial state $|\psi_0\rangle=|11\rangle$, which is also the ground state of our critical model at $B_z\gg1$. In this case
\begin{equation}
	|\langle\psi_0|U_q^\dagger\partial_{B_z}U_q|\psi_0\rangle|^2=s_I^2+s_z^2.
\end{equation}
Since
\begin{eqnarray}
    s_I&=&r_Iv_I+r_xv_x+r_zv_z\\
    &=&(\cos \Omega t,-\frac{h_x}{\Omega}\sin \Omega t,-\frac{h_z}{\Omega}\sin \Omega t)\cdot(-\frac{h_z}{\Omega}t\sin \Omega t,\\\notag
&&\frac{h_xh_z}{\Omega^3}\sin\Omega t-\frac{h_xh_z}{\Omega^2}t\cos \Omega t,-\frac{h_x^2}{\Omega^3}\sin \Omega t-\frac{h_z^2}{\Omega^2}t\cos \Omega t)\\\notag
&=&-\frac{h_z}{\Omega}t\sin \Omega t\cos \Omega t-\frac{h_x^2h_z}{\Omega^4}\sin^2\Omega t+\frac{h_x^2h_z}{\Omega^3}t\sin \Omega t\cos \Omega t\\\notag
&&+\frac{h_x^2h_z}{\Omega^4}\sin^2\Omega t+\frac{h_z^3}{\Omega^3}t\sin \Omega t\cos \Omega t\\\notag
&=&0\\\notag
s_z&=&r_Iv_z-r_zv_I\\\notag
&=&\cos \Omega t\times(-\frac{h_x^2}{\Omega^3}\sin \Omega t-\frac{h_z^2}{\Omega^2}t\cos \Omega t)-\frac{h_z}{\Omega}\sin\Omega t\times\frac{h_z}{\Omega}t\sin\Omega t\\\notag
&=&-\frac{h_x^2}{\Omega^3}\sin \Omega t\cos \Omega t-\frac{h_z^2}{\Omega^2}t\\\notag
s_z^2&=&\frac{h_x^4}{\Omega^6}\sin^2\Omega t\cos^2\Omega t+\frac{2h_x^2h_z^2}{\Omega^5}t\sin \Omega t\cos \Omega t+\frac{h_z^4}{\Omega^4}t^2,\notag
\end{eqnarray}
we have the QFI is
\begin{eqnarray}\label{FQ}
    \mathcal F(|\psi_\text{f}\rangle)&=&\frac{4h_z^2t^2}{\Omega^2}+\frac{4h_x^2}{\Omega^4}\sin^2\Omega t-4(\frac{h_x^4}{\Omega^6}\sin^2\Omega t\cos^2\Omega t+\frac{2h_x^2h_z^2}{\Omega^5}t\sin \Omega t\cos \Omega t+\frac{h_z^4}{\Omega^4}t^2\notag
)\\
&=&\frac{4h_x^2h_z^2}{\Omega^4}t^2+\frac{8h_x^2h_z^2}{\Omega^5}t\sin \Omega t\cos \Omega t+\frac{4h_x^4}{\Omega^6}\sin^4\Omega t+\frac{4h_x^2h_z^2}{\Omega^6}\sin^2\Omega t,
\end{eqnarray}
with the evolved state 
\begin{equation}\label{psif}
	|\psi_\text{f}\rangle=\left(\cos\Omega t-i\frac{h_z}{\Omega}\sin\Omega t\right)|a\rangle-i\frac{h_x}{\Omega}\sin\Omega t|b\rangle.
\end{equation}

In proximity of the critical point, we have $h_z\to0$ and the dominant term of the QFI is
\begin{equation}
	\mathcal F(|\psi_\text{f}\rangle)\sim\frac{4h_x^4}{\Omega^6}\sin^4\Omega t.
\end{equation}

\section{Optimal measurement and classical Fisher information}
The choice of measurement affect the precision. For a set of positive-operator valued measure (POVM) $\{\hat\Pi_m\}$ satisfying $\sum_m\hat\Pi_m=\mathds 1$ and quantum state $|\psi_f\rangle$ containing the information of $B_z$, we have the probability of measurement result $p(m|B_z)=\text{Tr}(|\psi_\text{f}\rangle\langle\psi_\text{f}|\hat\Pi_m)$. From $p(m|B_z)$ we can obtain the classcal Fisher information (CFI)
\begin{equation}\label{CFI}
	\mathcal{I}(\{\hat{\Pi}_m\})=\sum_i\frac{[\partial_{B_z} p(m|B_z)]^2}{p(m|B_z)},
\end{equation}
which bounds the attainable precision under POVM $\{\hat\Pi_m\}$ as
% The precision for estimating $B_z$ is
\begin{equation}
	\Delta B_z \geq \frac{1}{\nu \mathcal{I}(\{\hat{\Pi}_m\})} \geq \frac{1}{\nu \mathcal{F}}.
\end{equation}

To saturate the ultimate precision given by quantum Cram\'{e}r-Rao bound, optimal measurement is required such that
	$\mathcal{I}(\{\hat{\Pi}^\text{opt}_m\})=\mathcal F$. Here we construct the optimal measurement with the eigenstates of the symmetric logarithmic derivative (SLD) defined as \cite{Braunstein1994}
\begin{equation}
	L=2\left(\left|\partial_{B_z} \psi_\text{f}\right\rangle\left\langle\psi_f|+| \psi_\text{f}\right\rangle\left\langle\partial_{B_z} \psi_\text{f}\right|\right).
\end{equation}
For the encoded quantum state $|\psi_\text{f}\rangle$ in Eq. \eqref{psif}, we have
\begin{equation}
	L \sim\left(\begin{array}{cc}
2 r_I v_I+2 r_z v_z & r_x v_z+r_z v_x-i r_x v_I-i r_I v_x \\
r_x v_z+r_z v_x+i r_x v_I+i r_I v_x & 2 r_x v_x
\end{array}\right).
\end{equation}
At the critical point $B_z^c$, we have the optimal measurement  
\begin{equation}\label{optM}
	\{\hat\Pi_m^\text{opt}=|v_m^\text{opt}\rangle\langle v_m^\text{opt}|\}_{m=1}^4
\end{equation}
 with
\begin{equation}
	|v_1^\text{opt}\rangle=\frac{1}{2}\left(\begin{array}{c}
0 \\
1 \\
1 \\
\sqrt{2}
\end{array}\right), |v_2^\text{opt}\rangle=\frac{1}{2}\left(\begin{array}{c}
0 \\
1 \\
1 \\
-\sqrt{2}
\end{array}\right), |v_3^\text{opt}\rangle=\frac{1}{2}\left(\begin{array}{c}
\sqrt{2} \\
1 \\
-1 \\
0
\end{array}\right), |v_4^\text{opt}\rangle=\frac{1}{2}\left(\begin{array}{c}
-\sqrt{2} \\
1 \\
-1 \\
0
\end{array}\right).
\end{equation}

In the framework of local estimation, the parameter to be estimated is the smallest variations around a known parameter. In this case we can drive the critical system almost exactly to the critical point and perform the optimal measurement in Eq. \eqref{optM}. In our experimental demonstration, this is realized by applying a unitary operation $U_\text{tr}=\sum_m|v_m^\text{exp}\rangle\langle v^\text{opt}_m|$ before measurements with $\{|v_m^\text{exp}\rangle\}=\{|00\rangle,|01\rangle,|10\rangle,|11\rangle\}$ being the measurement basis of experimental apparatus. To assess the performance of parameter estimation, we extract $\mathcal{I}(\{\hat{\Pi}_m\})$ in Eq. \eqref{CFI} with the finite difference approach to obtain the susceptibility of probability, i.e., 
\begin{equation}
	\partial_{B_z} p(m|B_z)\approx \frac{p(m|B_z+\delta)-p(m|B_z-\delta)}{2\delta},
\end{equation}
where $\delta$ is set to 0.06 in our experiment.

\section{Comparison with the equilibrium approach}
Criticality-enhanced quantum sensing via the equilibrium approach utilizes the ground state in the adiabatic process or the steady state in a driven-dissipative process. In our non-equilibrium approach, we focus on the unitary quantum quench process of a close quantum system. For comparison, we also consider the unitary adiabatic passage for the equilibrium approach. 

The adiabatic passage can be described by a time-dependent process
\begin{equation}
	\mathcal H[A(s)]=[1-A(s)]\mathcal H_0+A(s)\mathcal H_\text{f},
\end{equation}
where $\mathcal H_0=\widetilde H(B_{z0}),\mathcal H_\text{f}=\widetilde H(B_{z\text{f}})$, $A(s)$ is the adiabatic path with $s=t/T\in[0,1]$ being the normalized time. The design of $A(s)$ is crucial for the precision scaling. To achieve the Heisenberg scaling, we design $A(s)$ using a numerical approach. We start from $A_0=0$. For the $j$-th step, $A_j$ is the mimimal value such that $1-|\langle g(A_j)|e^{-i\mathcal H(A_j)\Delta t}|g(A_{j-1})\rangle|^2\ge P_c$, where $|g(A_j)\rangle$ denotes the ground state of $\mathcal H(A_j)$, $P_c\ll1$ is the threshold of the fidelity decay. Suppose $A_j>1$ when $j=N$, we terminate the loop and obtain the adiabatic path as $A(j/N)=A_j$. Here $\Delta t$ is fixed and $P_c=10^{-4}$ to guarantee the evolved state is close to its ground state.

\begin{figure*}
		\begin{center}
			\includegraphics[scale=1.6]{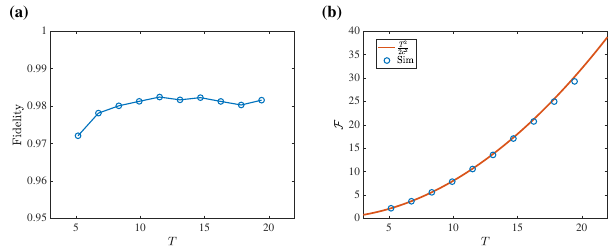} 
			\caption{(a) The fidelities of adiabatically evolved states for $c=2.5$, all of which are larger than 97.2\%. (b) The QFI of adiabatically evolved states. It conforms well with the theoretical $T^2/2c^2$.}\label{FigS3}
		\end{center}
\end{figure*}

In the design of the numerical adiabatic passage, we gradually increase $A_n$ with a stepsize $\Delta A$ to search the minimal one such that $1-|\langle g(A_j)|e^{-i\mathcal H(A_{j-1}+n_j\Delta A)\Delta t}|g(A_{j-1})\rangle|^2\ge P_c$. This suggests $1-|\langle g(A_j)|e^{-i\mathcal H(A_j)\Delta t}|g(A_{j-1})\rangle|^2=P_c$ for a sufficiently small $\Delta A$. Since $P_c\ll1$, $A_j$ is close to $A_{j-1}$. According to the perturbation theory, we have the evolved state under $\mathcal H(A_j)$
\begin{equation}
	P_c=\frac{n_{j}^2 \Delta A^2}{2} \sum_{k \neq g} \frac{|\left\langle k\left|\mathcal{H}_\text{f}-\mathcal{H}_0\right| g\right\rangle|^2}{\left(E_k-E_g\right)^2},
\end{equation}
where $\{|k\rangle\}$ denote the excited states of $\mathcal H(A_{j-1})$. We thus have $n_j\Delta A=\sqrt{P_cf(A_{j-1})}$ with $f(A_{j-1})=2/\left(\sum_{k \neq g} \frac{|\left\langle k\left|\mathcal{H}_\text{f}-\mathcal{H}_0\right| g\right\rangle|^2}{\left(E_k-E_g\right)^2}\right)$. The time consumption of adiabatic passage is given by \cite{jansen2007,Albash2018}
\begin{equation}
	T \gg \max _{s \in[0,1]} \frac{\left|\left\langle k\left|\partial_s \mathcal{H}[A(s)]\right| g\right\rangle\right|}{\left(E_k-E_g\right)^2}, \forall k \geq 1.
\end{equation} 
Since
\begin{equation}
	\begin{aligned}
\left.\partial_s \mathcal{H}[A(s)]\right|_{s=\frac{j-1}{N}} & \approx \frac{\mathcal{H}\left(A_{i+1}\right)-\mathcal{H}\left(A_i\right)}{\Delta s} \\
& =\frac{\sqrt{P_c f\left(A_{j-1}\right)}\left(\mathcal{H}_{\mathrm{f}}-\mathcal{H}_0\right)}{\Delta s},
\end{aligned}
\end{equation}
we have 
\begin{equation}
	\frac{\left|\left\langle k\left|\partial_s \mathcal{H}[A(s)]\right| g\right\rangle\right|}{\left(E_k-E_g\right)^2} \leq \frac{\sqrt{2P_c}}{\Delta s} \frac{1}{\left(E_1-E_g\right)}.
\end{equation}
Here we truncate to the first excited state with the eigenvalue $E_1$ and have the time required $T=c/B_x$, where $c$ is a constant determining the speed of adiabatic passage. Based on the QFI at the critical point, we have $\mathcal F\sim T^2$, thus a Heisenberg scaling.

The value of $c$ can be determined through numerical simulations. It should not be too small, as this would result in a low fidelity between the adiabatically evolved state and the ground state. Conversely, it should not be too large, as this would lead to an unnecessary waste of time resources. In our simulation, we set $c=2.5$. As shown in Fig.~\ref{FigS3}(a), in this case, the fidelities of adiabatically evolved states are larger than 97.2\%. The corresponding QFI of evolved states is shown in Fig.~\ref{FigS3}(b), which conforms well with the theoretical $T^2/2c^2$.
\section{Scaling of the critical region for sensing}
In general, the significantly enhanced precision at the critical point is accompanied by a rapid decay in precision away from the critical point. This leads to a narrow critical region for sensing and poses a great challenge to the application of critical quantum metrology in global sensing, in which the parameter to be estimated varies over a wide range. Based on the region with criticality-enhanced QFI and maximum distinguishable signal range \cite{Degen2017}, we provide a detailed analysis of the critical region for sensing in our critical model.

\begin{figure*}
		\begin{center}
			\includegraphics[scale=1.6]{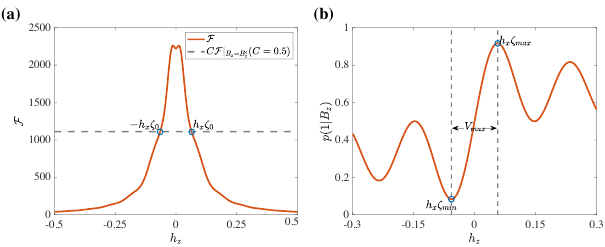} 
			\caption{(a) The critical region for sensing. When $h_z\in(-h_x\zeta_0,h_x\zeta_0)$, the QFI does not significantly decay below the criticality-enhanced value multiplied by a constant $C$. The length of this region scales as $\tau^{-1}$ or $(\mathcal F|_{B_z=B_z^c})^{-\frac{1}{2}}$. (b) The maximum distinguishable signal range in our critical sensing protocol. When $h_z\in(h_x\zeta_{min},h_x\zeta_{max})$, there is a one-to-one correspondence between $B_z$ and $p(i|B_z)$, and $B_z$ can thus be uniquely distinguished from each measured probability. This range also shrinks with a scaling $\tau^{-1}$ or $(\mathcal F|_{B_z=B_z^c})^{-\frac{1}{2}}$.}\label{FigS2}
		\end{center}
\end{figure*}

\begin{itemize}
	\item \textbf{The region with criticality-enhanced QFI}

	We define the region with criticality-enhanced QFI as the range of $B_z$ in which the QFI does not significantly decay below the maximum (exactly at the critical point) multiplied by a constant $C\in(0,1)$, that is
	 \begin{equation}\label{region}
	 	\{B_z:\mathcal F|_{B_z}\ge C\mathcal F|
    	_{B_z=B_z^c}\ \text{with the constant}\ C\in(0,1)\}.
	 \end{equation}
	 For $C=0.5$, this definition can be visually illustrated in Fig.~\ref{FigS2}(a).
	 
	 In the following, we analyze how this region shrinks when the QFI at the critical point increases. 
	 
	 We can decide the width of the region in Eq. \eqref{region} from the equation
	 \begin{equation}\label{zero_region}
	 	\mathcal F|_{B_z}= C\mathcal F|_{B_z=B_z^c}.
	 \end{equation}
	 By substituting Eq. \eqref{FQ} and our protocol duration $\tau=\pi/\Delta_c$ ($\Delta_c = 2\sqrt{2}B_x$ represents the corresponding energy gap) into Eq. \eqref{zero_region}, we have an equation about $h_z$, i.e.,
	 \begin{equation}
	 	\frac{\pi^2h_x^2h_z^2}{\Omega^4}+\frac{\pi h_x^3h_z^2}{\Omega^5}\sin\frac{\pi\Omega}{2h_x}\cos \frac{\pi\Omega}{2h_x}+\frac{4h_x^6}{\Omega^6}\sin^4\frac{\pi\Omega}{2h_x}+\frac{4h_x^4h_z^2}{\Omega^6}\sin^2\frac{\pi\Omega}{2h_x}-4C=0.
	 \end{equation}
	 In fact, we do not need the exact solution of this equation but just rewrite it as
	 \begin{equation}\label{equ_region}
	 	\frac{\pi^2\zeta^2}{(1+\zeta^2)^2}+\frac{\pi \zeta^2}{2(1+\zeta^2)^{\frac{5}{2}}}\sin\pi\sqrt{1+\zeta^2}+\frac{4}{(1+\zeta^2)^3}\sin^4\frac{\pi\sqrt{1+\zeta^2}}{2}+\frac{4\zeta^2}{(1+\zeta^2)^3}\sin^2\frac{\pi\sqrt{1+\zeta^2}}{2}-4C=0,
	 \end{equation}
	 where $\zeta=h_z/h_x$. Suppose the two solutions of Eq. \eqref{equ_region} is $\zeta_{\pm}=\pm\zeta_0$ with $\zeta_0>0$, we have the width of the critical region for sensing 
	 \begin{equation}\label{Lregion}
	 	L=2h_x\zeta_0.
	 \end{equation}
	 For a chosen constant $C$, $\zeta_0$ is decided. Moreover, in our protocol, we work in a regime $h_x\propto \tau^{-1}$ or $\propto (\mathcal F|_{B_z=B_z^c})^{-1/2}$. Consequently, Eq. \eqref{Lregion} indicates that the length of the critical region for sensing shrinks with the protocol duration and the precision at critical point.
	 \item\textbf{Maximum distinguishable signal range}
	 
	 For a measured probability $p(m|B_z)$, there can be more than one parameter to be estimated $B_z$ that corresponds to this value of $p(m|B_z)$. Consequently, a unique assignment requires a priori knowledge of $B_z$ that lies in a range, in which there is a one-to-one correspondence between $B_z$ and $p(m|B_z)$. This defines a maximum distinguishable range of $B_z$, denoted as $V_\text{max}$ and illustrated in Fig.~\ref{FigS2}(b). As an easy example, the maximum distinguishable phase range in a standard Ramsey interferometry is $2\pi$. In our protocol, we specifically fix the measurement as $\{\hat\Pi_m^\text{opt}\}_{m=1}^4$ in Eq. \eqref{optM} and measure the evolved states. The probability on $|v_1^\text{opt}\rangle$ is 
	 \begin{equation}\label{probabi}
	 	p(1|B_z)=\frac{1}{2}+\frac{h_xh_z}{\Omega^2}\sin^2\frac{\pi\Omega}{2h_x}.
	 \end{equation}
	 As we can see from Fig.~\ref{FigS2}(b), $V_\text{max}$ is the length between two neighboring extreme point. Similar to Eq. \eqref{equ_region}, we can rewrite Eq. \eqref{probabi} as
	 \begin{equation}\label{probabi}
	 	p(1|B_z)=\frac{1}{2}+\frac{\zeta}{1+\zeta^2}\sin^2\frac{\pi}{2}\sqrt{1+\zeta^2}.
	 \end{equation}
	 Suppose the two extreme points in Fig.~\ref{FigS2}(b) are $\zeta_\text{min}$ and $\zeta_\text{max}$, we thus have
	 \begin{equation}\label{Vmax}
	 	V_\text{max}=h_x\left|\zeta_\text{min}-\zeta_\text{max}\right|.
	 \end{equation}
	 Similar to Eq. \eqref{Lregion}, Eq. \eqref{Vmax} means that the maximum distinguishable signal range also shrinks with the protocol duration and the precision at the critical point.
\end{itemize}

\section{Procedure of experimental adaptive parameter estimation}
To fully exploit the criticality-enhanced precision while enable the application of critical system in global sensing, we adopt an adaptive strategy by incorporating feedback control into Bayesian estimation \cite{Bayes1763}. The crucial distinction from standard Bayesian estimation lie in the introduction of an additional control field that is updated according to each measurement and drives the quantum system close to its critical point. The experimental proceduce for the adaptive parameter estimation protocol is elaborated with a pseudocode in Algorithm \ref{Bayes}.

\begin{algorithm}[H]
\caption{Bayesian estimation incorporated with feedback control}
\label{Bayes}
\begin{algorithmic}[1]
\Statex \textbf{Input:} Initial state $\rho_0=|\psi_0\rangle\langle\psi_0|$
\Statex \qquad\quad The range of $B_z$, i.e., $B_z\in[B_z^\text{min},B_z^\text{max}]$
\Statex \qquad\quad An initial prior distribution $p_0(B_z)$
\Statex \qquad\quad Maximal number of experimental measurements $\nu$
\Statex \textbf{Output:} The posterior distribution $p\left(B_z \mid \vec{m}_\nu\right)$ updated according to $\nu$ measurement results $\vec{m}_\nu$
\State Set $k=1$, $p(B_z|m_0)=p_0(B_z)$, initial estimate $\hat B_z^0=\int B_z p_0(B_z)dB_z$, initial control field $B_z^{\text{ctrl,}0}=B_z^c-\hat B_z^0$
\While{$k\le\nu$}
	\State Encoding: $\rho(B_z^{\text{r}}+B_z^{\text{ctrl,}k})=U\rho_0U^\dagger$, where $U=\exp[-i {\widetilde{H}(B_x,B_z^{\text{r}}+B_z^{\text{ctrl,}k})}\tau],\tau=\pi/\Delta_c$, and $B_z^{\text{r}}$ is the assumed true value of $B_z$
    \State Measurement: Perform $\mathcal M^\text{opt}=\{|v_m^\text{opt}\rangle\langle v_m^\text{opt}|\}$ on $\rho(B_z^{\text{r}}+B_z^{\text{ctrl,}k})$ and record the outcome $m_k$ with $\vec{m}_k=\{m_1,m_2,\dots,m_k\}_{m=1}^4$
    \State Calculate the conditional probability $p(m_k|B_z+B_z^{\text{ctrl,}k})=\text{Tr}[\rho(B_z+B_z^{\text{ctrl,}k})|v_m^\text{opt}\rangle\langle v_m^\text{opt}|]$
    \State Update the posterior distribution: $p(B_z|\vec{m}_{k})\leftarrow \mathcal Np(m_k|B_z+B_z^{\text{ctrl,}k})p(B_z|\vec{m}_{k-1})$ with $\mathcal N=[\int p(m_k|B_z+B_z^{\text{ctrl,}k})p(B_z|\vec{m}_{k-1})]^{-1}$.    
    \State Update the estimation: $\hat B_z^k\leftarrow\int B_z p(B_z|\vec{m}_k)dB_z$
    \State Update the control field: $B_z^{\text{ctrl,}k+1}\leftarrow B_z^c-\hat B_z^k$
    \State Evaluation of precision: $\Delta^2\hat B_z=\int p(B_z|\vec{m}_{k})(B_z-\hat B_z^k)^2dB_z$
    \State $k\leftarrow k+1$
\EndWhile
\end{algorithmic}
\end{algorithm}
\section{Robustness against noise}
\begin{figure*}
		\begin{center}
			\includegraphics[scale=1.6]{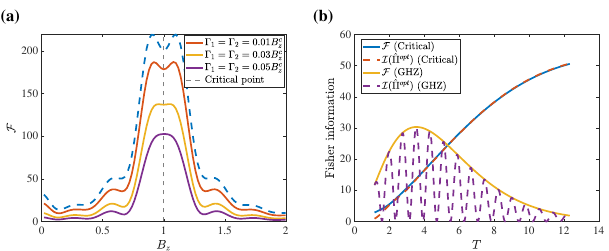} 
			\caption{The performance of our critical sensing scheme under Markovian noise. (a) plots the QFI as a function of $B_z$ under noise, where $B_x=0.1,\tau=\pi/\Delta_c$. (b) plots the QFI and the CFI under optimal measurement at the critical point $B_z^c$, in which we set $\Gamma_1=\Gamma_2=0.1B_z^c$. For comparison, the performance of the Ramsey sequence with the optimal GHZ state is also depicted.}\label{FigS4}
		\end{center}
\end{figure*}
We investigate the robustness of our criticality-enhanced quantum sensing scheme under Markovian noise. The evolution of quantum system can be described by the following quantum master equation
\begin{eqnarray}
	\dot{\rho}&=&-i[\widetilde H(B_z),\rho]+\mathcal L\rho,\\
	\mathcal L\rho&=&\Gamma_1(\mathcal D[\sigma_z^1]+\mathcal D[\sigma_z^2])+\Gamma_2(\mathcal D[\sigma_-^1]+\mathcal D[\sigma_-^2]), 
\end{eqnarray}
where $\mathcal{D}[\hat{A}]=\hat{A} \rho \hat{A}^{\dagger}-\hat{A}^{\dagger} \hat{A} \rho / 2-\rho \hat{A}^{\dagger} \hat{A} / 2$, $\Gamma_{1,2}$ denote the rate of dephasing and decay, respectively.

 We first study the advantage of the critical point under noise. The numerical results of QFI as a function of $B_z$ under different rates of Markovian noise are shown in Fig.~\ref{FigS4}(a). Though the noise degrades the metrological performance, the advantage of the critical point is still evident. We then investigate the precision at the critical point as a function of time consumption $T$ under noise. Apart from QFI, we assess the metrological performance via CFI under optimal measurement. For comparison, we also depict the performance of the Ramsey sequence with the optimal GHZ state $|GHZ\rangle=(|00\rangle+|11\rangle\rangle)/\sqrt2$, in which the encoding dynamics is $H_\text{Ram}=B_z(S_z^1+S_z^2)$. It can be seen in Fig.~\ref{FigS4}(b) that the QFI of the Ramsey sequence grows faster than our critical approach, but rapidly decays when $T$ increases. Moreover, due to the effect of noise, the performance of the original optimal measurement is no longer able to saturate the QCRB, which leads to an oscillating CFI. This result aligns with previous works \cite{Liu2017,Zhang2022}. For our critical sensing scheme, the QFI decays more slowly and finally surpasses the GHZ state. Meanwhile, the optimal QCRB is always saturated by the original optimal measurement. 

\section{Device design and experimental setup}
The superconducting quantum processor used in our experiment is fabricated using flip-chip technology. The processor consists of a qubit chip and a carrier chip. The size of the qubit chip and carrier chip are $45\times39$ $\mu m$ and $20\times20$ $\mu m$, respectively. The two chips are separated by an indium bump stopper which can control the spacing between the two chips. The distance between the qubit chip and carrier chip is $10\pm 1$ $\mu m$. The carrier chip contains transmission lines, Purcell filters, and control lines for the qubits and couplers. The qubit chip includes qubits, couplers, and readout resonators. All the lines in the carrier chip are covered with aluminum air bridges, isolating the carrier from coupling with the qubits and couplers, with air bridges absent only where coupling is required. All qubits and couplers are of the floating pad transmon,  coupled through capacitance, and the effective coupling strength between qubits can be controlled by adjusting the frequency of the couplers. The qubit chip does not contain control lines, reducing multiple micro-fabrication process steps, allowing the qubits to maintain high coherence performance. The $E_\text{C}$ of the qubits and couplers are 189 MHz and 153 MHz, respectively. 
The Kappa of the readout resonator is about 2 MHz, and the dispersive shift of the readout resonator between qubit states $|0\rangle$ and $|1\rangle$ is approximately 0.51 MHz.

\begin{figure}[h]
	\begin{center}
		\includegraphics[scale=0.5]{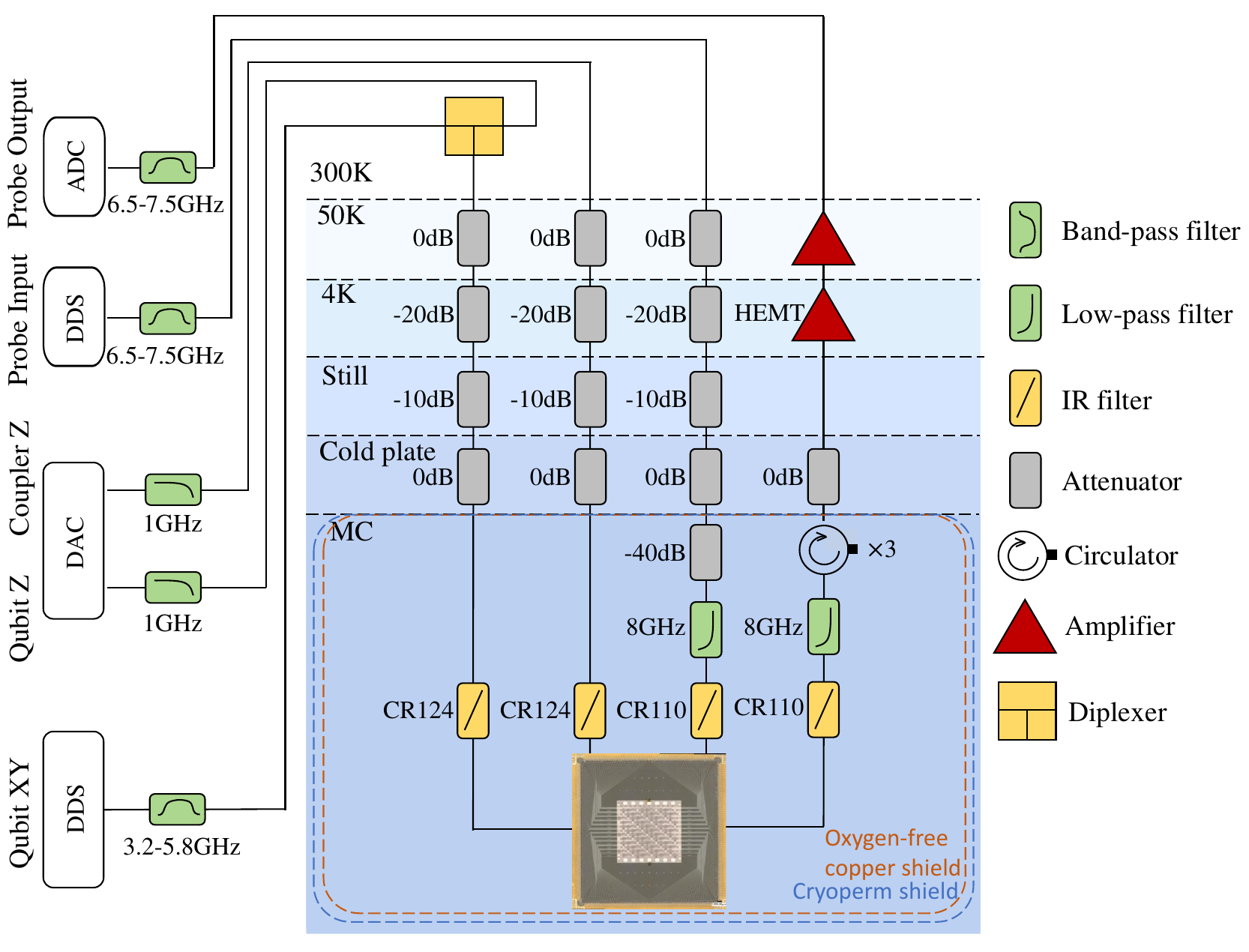} 
		\caption{A schematic of the experimental setup.}\label{FigS5}
	\end{center}
\end{figure}

Our experimental setup is schematically depicted in Fig.~\ref{FigS5}, which encompasses the cryogenic system, associated wirings, microwave components, and room-temperature measurement electronics.

The qubit drive (XY) signals and the probe input signals are produced by a direct digital synthesis (DDS) which can support parameterized waveforms with a sampling rate of up to 10 GS/s and a vertical resolution of 14 bits. The probe output signals are digitized by an analog-to-digital converter (ADC) which perform real-time FPGA demodulation with a sampling rate of up to 4 GS/s and a vertical resolution of 14 bits. The qubits and couplers flux bias (Z) control signals are generated from a digital-to-analog converter (DAC) with a sampling rate of up to 2 GS/s and a vertical resolution of 16 bits.

The XY and fast Z control signals of each qubit are combined by a diplexer at room temperature, pass through an attenuated cryogenic coaxial cable, and are filtered by an 8 GHz low-pass filter and a CR124 infrared (IR) filter before reaching the device at the 10-20 mK mixing-chamber (MXC) stage of the Bluefors dilution refrigerator.  In order to mitigate the impact of spurious radiation and flux noise, a light-tight oxygen-free copper shield and a cryoperm shield are added outside the device. 

The probe input signal undergoes a total attenuation of 70 dB and is filtered by an 8 GHz low-pass filter and a CR110 IR filter, ensuring the suppression of thermal photon noise. The probe output signal sequentially passes an 8 GHz low-pass filter, a CR110 IR filter, and three cryogenic circulators. It is subsequently amplified by a high electron mobility transistor (HEMT) amplifier at the 4 K stage and another amplifier at the 50 K stage before ultimately being captured by the measurement equipment.

\section{Characterization and calibration}
Upon cooling down, we perform a series of fundamental characterizations and calibrations, including Z pulse distortion correction, timing calibration, single-qubit gate calibration, readout optimization, and controlled-Z (CZ) gate calibration. As a result of these calibrations, the main parameters of the two qubits we used are summarized in Table \ref{device parameters}, providing a comprehensive overview of this quantum device. 

\begin{table}[H]
	\centering
	\begin{tabular}{ m{1.3cm} m{1.3cm} m{1.3cm} m{1.3cm} m{1.3cm} m{1.3cm} m{1.3cm} m{1.3cm} m{1.3cm} m{1.3cm} m{1.3cm} m{1.3cm} } 
		\toprule
		Qubit & $\omega_r/2\pi$ & $\omega_{\text{max}}/2\pi$ & $\omega_{\text{idle}}/2\pi$ & $\alpha$ & $T_1$ & $T_2^*$ & $T_2^{\text{echo}}$ & $F_\text{g}$ & $F_\text{e}$ & $F_\text{1Q}$ & $F_\text{CZ}$\\ 
		& (GHz) & (GHz) & (GHz) & (MHz) & ($\mu s$) & ($\mu s$) & ($\mu s$) & ($\%$) & ($\%$) & ($\%$) & ($\%$)\\
		\midrule
		$\text{Q}_1$ & 7.226 & 4.420 & 4.414 & -193 & 97.6 & 48.3 & 78.4 & 95.9 & 94.6 & 99.92\\
		&&&&&&&&&&&99.1\\
		$\text{Q}_2$ & 6.846 & 4.294 & 4.285 & -185 & 86.4 & 45.3 & 74.1 & 97.5 & 94.6 & 99.92\\
		\bottomrule
	\end{tabular}
	\caption{Main parameters of the two qubits used in this work. $\omega_r$ is the readout resonator frequency. $\omega_\text{max}$ and $\omega_\text{idle}$ are the qubit maximum and idle frequencies. $\alpha$ is the qubit anharmonicity. $T_1$, $T_2^*$, and $T_2^\text{echo}$ are the qubit energy relaxation time, Ramsey decay time, and spin-echo decay time at the idle frequency, respectively (see Fig.~\ref{FigS6}). $F_\text{g}$ and $F_\text{e}$ represent the simultaneous readout fidelities of the ground and first excited states. $F_\text{1Q}$ and $F_\text{CZ}$ denote the single-qubit gate and CZ gate fidelities, both of which are determined via the Clifford-based randomized benchmarking (RB) \cite{Magesan2012RB,Knill2008RB}.}
	\label{device parameters}
\end{table}

\begin{figure}[h]
	\begin{center}
		\includegraphics[scale=0.62]{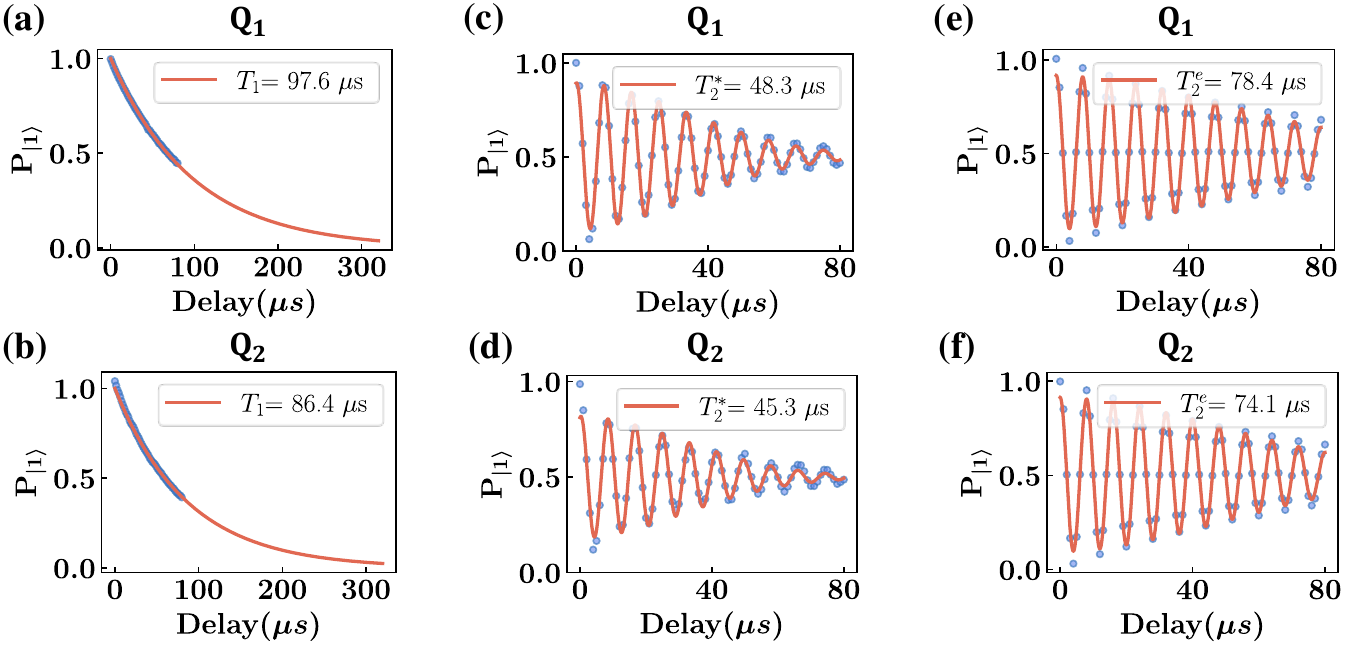} 
		\caption{(a-b) The relaxation times, (c-d) the Ramsey decay times, and (e-f) the spin-echo decay times of the two qubits.}\label{FigS6}
	\end{center}
\end{figure}

\begin{itemize}
	\item \textbf{Z pulse distortion correction}
	
	The shape of the Z control pulse is distorted as it passes through various electronic components. Such distortion can induce deviations in the qubit frequency, leading to unintended phase errors and a subsequent decline in gate fidelity. To correct the Z pulse distortion, we follow the distortion detection scheme in Ref.~\cite{bao2022fluxonium}.  As illustrated in the inset of Fig.~\ref{FigS7}(a), after a very long square wave with a large enough amplitude $v_{0}$, a Ramsey-like pulse sequence with a 20 ns probe Z pulse sandwiched by two $\pi$/2 pulses is applied to probe the accumulated phase. We calculate the corresponding trailing amplitude $v_{d}(t)$ with the phase via the mapping relationship obtained by varying the amplitude of the probe Z pulse. The trailing amplitude is then fitted by a multi-component exponential function, $v_{d}(t)/v_0=\sum_i a_ie^{-t/\tau_i}$. With the fit parameters, we pre-distort the input signal, effectively rectifying the Z pulse distortion.
	
	\item \textbf{Timing calibration}
	
	Given the inherent disparities in the lengths of transmission lines and the distinct response times of different instruments, the temporal arrival of signals at qubits through different channels varies. In our experiment, we take the qubit XY control as a reference. Utilizing the sequences illustrated in Fig.~\ref{FigS7}(b-c), we measure the delay between the XY control and Z control of individual qubits, as well as between the XY control of both qubits and the coupler Z control. To align the timing between the qubit XY and Z control channels,  we apply an X gate on the XY control and a square wave of equivalent duration on the Z control. When the two overlap, the X gate cannot fully excite the qubit. By altering the position of the square wave, we monitor the qubit population as a function of the delay. The measurement of the delay between the qubit XY control and the coupler Z control is carried out in a similar way. Timing calibration is achieved by manually reordering the sequence of different channels to compensate for the measured delays.
	
	\begin{figure}[h]
		\begin{center}
			\includegraphics[scale=0.63]{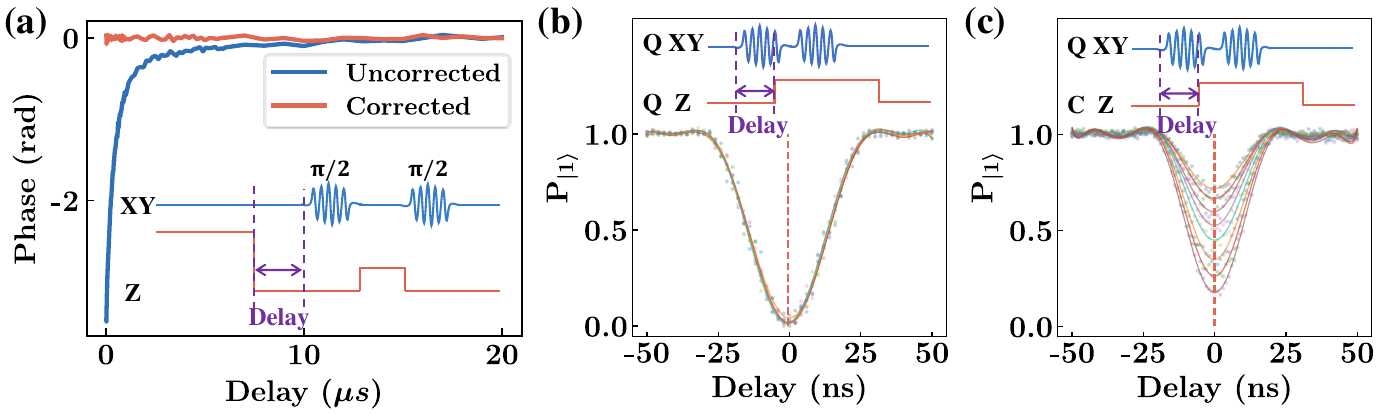} 
			\caption{(a) The correction of a distorted Z control pulse. The blue (red) line shows the unwrapped qubit phase shift resulting from a Z pulse without (with) predistortion.  (b) The timing alignment between qubit XY and Z controls. (c) The timing alignment between the qubit XY control and the coupler Z control. Different colored dots represent the results corresponding to different Z pulse amplitudes, solid lines are the fits to them, and the dashed line is an average of the bottoms of these fitted curves. Pulse sequences are illustrated in the insets.}\label{FigS7}
		\end{center}
	\end{figure}
	
	\item \textbf{Single-qubit gate calibration}
	
	To maximize the coherence time and avoid unintended couplings, the idle point of each qubit is selected to be in proximity to its sweet point, while simultaneously maintaining a safe distance from the frequencies of other undesired two-level systems. The remaining unused nearest and next nearest neighbors are tuned to idle frequencies as far away as possible from these two qubits. After fine calibration of qubit frequency with the Ramsey method, we turn off the ZZ coupling by tuning the coupler frequency. We enhance the performance of the single-qubit gate by alternating the calibration of the $\pi$ pulse amplitude and the optimization of the derivative removal by adiabatic gate (DRAG) parameters. Subsequently, we optimize the frequency and amplitude of the readout microwave pulse. Consequently, Fig.~\ref{FigS8} presents the simultaneous readout signals for both the ground and excited states of the two qubits. We employ RB to characterize the fidelity of our single-qubit gates, achieving results that demonstrate a fidelity exceeding $99.9\%$.
	
	For each data point depicted in Fig. 2(c) and Fig. 3(b) of the main text, we conduct ongoing monitoring and calibration of the single-qubit gates and readout. Additionally, we assemble the measurement calibration matrix, denoted as $M$, by preparing each state within the two-qubit Hilbert space and recording the corresponding probabilities. We then compute the vector of error-mitigated measurement outcomes using the expression $V = M^{-1}V_{raw}$, where $V_{raw}$ represents the vector of raw measurement outcomes. 
	\begin{figure}[h]
		\begin{center}
			\includegraphics[scale=0.66]{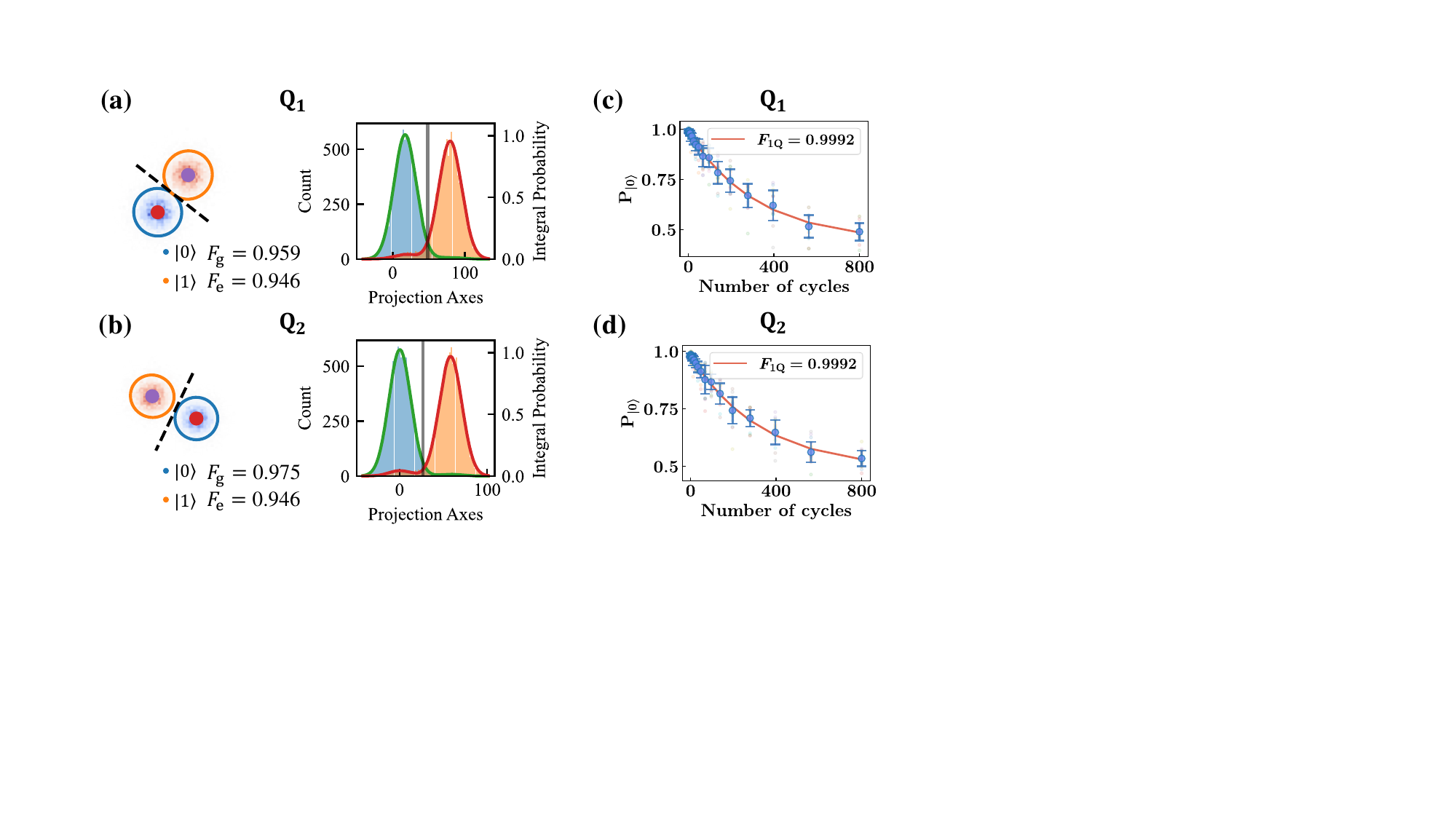} 
			\caption{(a-b) Left: scatter diagram of the readout signals for the two qubits in the I/Q plane. Right: Distribution of projections onto the lines connecting their centers.  (c-d) RB of single-qubit gates.  }\label{FigS8}
		\end{center}
	\end{figure}

	\item \textbf{CZ gate calibration}
	
	The quantum quench evolution in our experiment is decomposed into an equivalent quantum circuit consisting of single-qubit and CZ gates. To realize the CZ gate, we fix the gate length to 80 ns and vary the amplitudes of the square wave on the Z control of $\text{Q}_2$ and the hyperbolic cosine pulse on the Z control of the coupler, denoted as $amp_{\text{Q}_2}$ and $amp_\text{C}$. Utilizing the pulse sequences depicted in Fig.~\ref{FigS9}, we measure the conditional phase $\phi$ and state population leakage \cite{Foxen2020CZ}. The data points that exhibit a conditional phase near $\pi$ and minimal leakage are selected, respectively. Their intersection serves as an initial working point, as highlighted by the green triangle. We then scan $amp_{\text{Q}_2}$ around this initial value and replicate the CZ gate n cycles to amplify the gate error, thus realizing a fine calibration of gate parameters. In addition, we compensate for the phase accumulated on individual qubits during the CZ gate operation by applying virtual-Z gates. Following this calibration process, we achieve a CZ gate with an error rate of less than 1$\%$, as confirmed by RB.
	\begin{figure}[h]
		\begin{center}
			\includegraphics[scale=0.7]{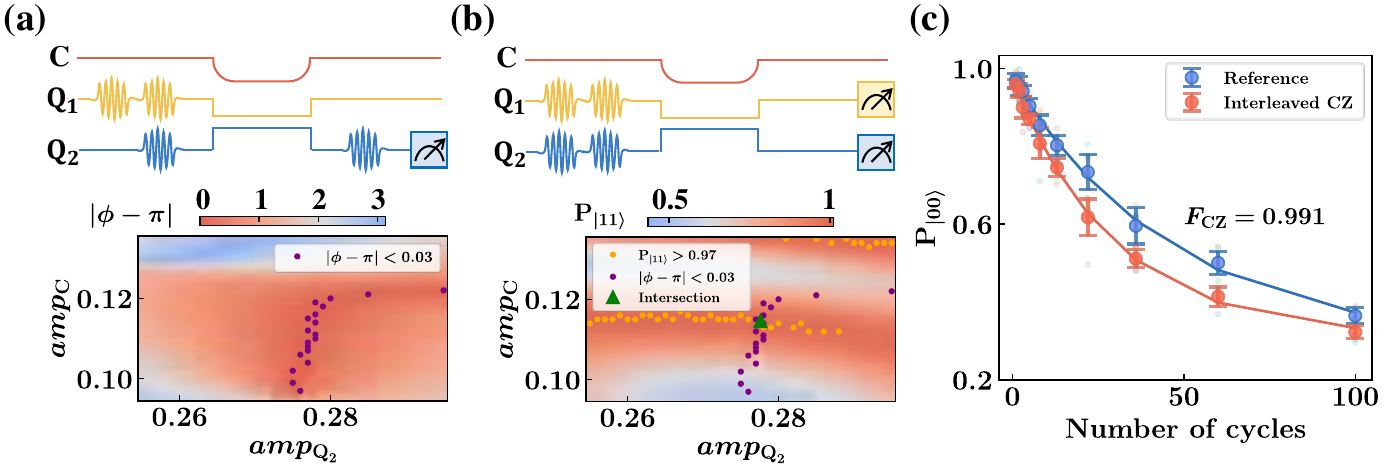} 
			\caption{(a) The pulse sequence for the measurement of conditional phase $\phi$ and the corresponding result. After applying a X or I gate on $\text{Q}_1$, we perform a Ramsey-like sequence on $\text{Q}_2$ to capture the phase shift. Purple dots represent the data points where $|\phi-\pi|<0.03$. (b) The pulse sequence for the measurement of population leakage and the corresponding result. The initial state is prepared in $|11\rangle$. The population $P_{|11\rangle}$ is measured after implementing the CZ gate. Orange dots represent the data points where $P_{|11\rangle}>0.97$. The intersection, indicated by the green triangle, shows the initial value of $amp_\text{C}$ and $amp_{\text{Q}_2}$. (c) RB of the CZ gate. }\label{FigS9}
		\end{center}
	\end{figure}

\end{itemize}

\section{Quantum circuit}

\begin{figure}[h]
	\begin{center}
		\includegraphics[scale=0.56]{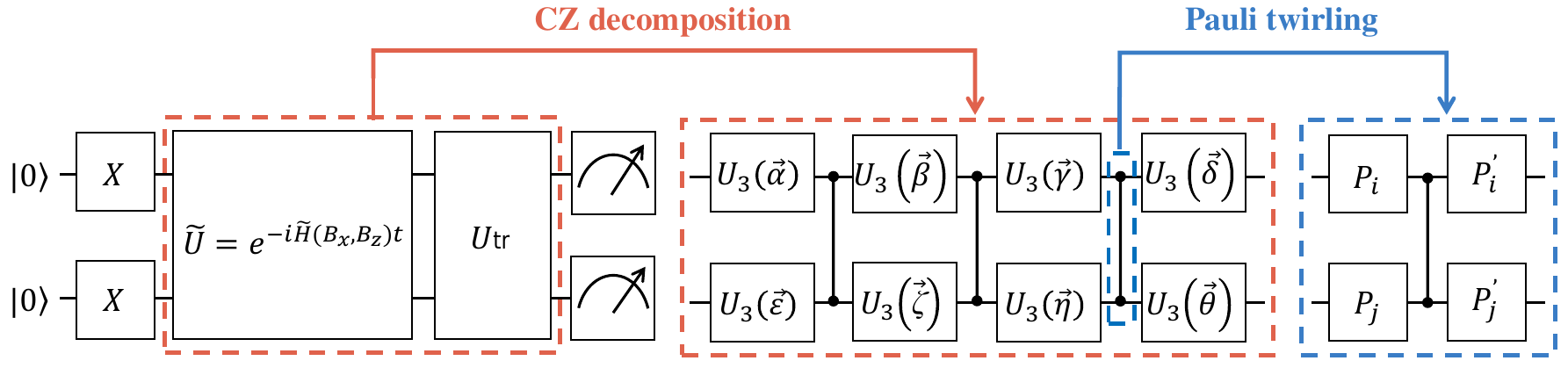} 
		\caption{Circuit decomposition and Pauli twirling scheme.}\label{FigS10}
	\end{center}
\end{figure}

Any two-qubit unitary operation can be decomposed into three CNOT gates and eight $U_3$ gates. For simplicity and accuracy, all the quench evolutions represented by $\tilde{U} = e^{-i\tilde{H}(B_x,B_z)t}$ in our experiments are decomposed in this way and then transpiled to the hardware native CZ gates. In addition, as for the experiments that incorporate optimal measurement, the unitary transformation $U_\text{tr}$ is combined with $\tilde{U}$ and then decomposed together into three CZ gates and eight $U_3$ gates (see Fig.~\ref{FigS10}). Each $U_3$ gate is decomposed into $R_z(\phi_1)R_x(-\pi/2)R_z(\phi_0)R_x(\pi/2)R_z(\phi_2)$, where $R_z(\phi_i)$ are virtual-Z gates.

To further improve the circuit fidelity, we employ Pauli twirling \cite{Wallman2016PT}, an effective technique to mitigate coherent errors by converting an arbitrary noise channel into a stochastic Pauli error channel. It is implemented by sandwiching the two-qubit gates, which are CZ gates in our experiments, with additional twirling gates randomly drawn from a set of Pauli operations $\{I, X, Y, Z\}^{\otimes N}$ for N qubits, ensuring that the net operation remains unchanged. Each twirling gate can be integrated with the adjacent original $U_3$ gates and recompiled into a new $U_3^{\prime}$ to prevent additional circuit cost. 

\section{Experimental error analysis}

\begin{figure}[h]
	\begin{center}
		\includegraphics[scale=0.5]{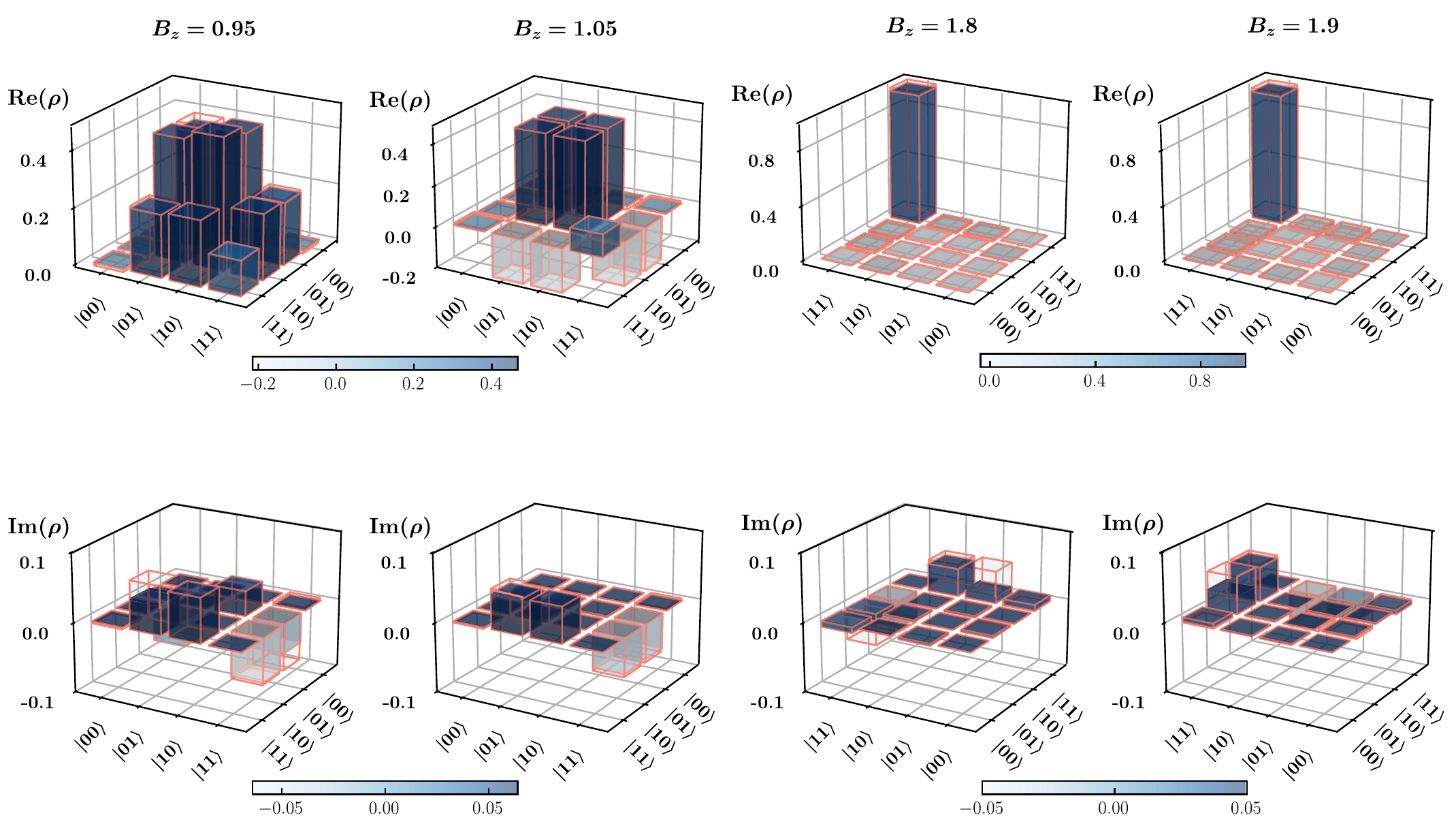} 
		\caption{Experimentally reconstructed density matrices. The solid blue bars and red wireframes are the measured and ideal quantities, respectively. }\label{FigS11}
	\end{center}
\end{figure}

\begin{figure}[h]
	\begin{center}
		\includegraphics[scale=0.75]{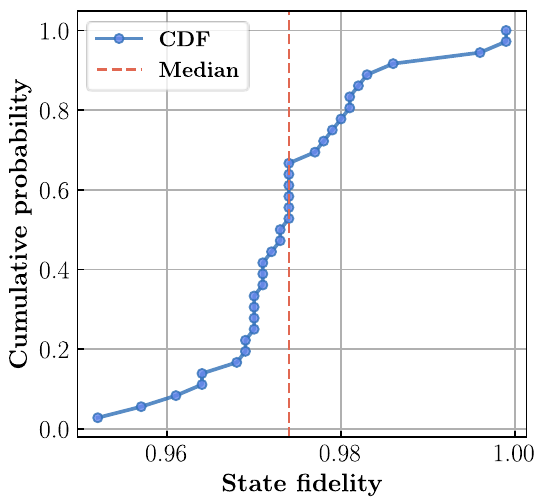} 
		\caption{Cumulative probability distribution of the state fidelities. The red line shows the median.
		}\label{FigS12}
	\end{center}
\end{figure}

To obtain Fig.2(c) in the main text, we extract the QFI via reconstructing the density matrices of the quenched states by quantum state tomography. In Fig.~\ref{FigS11}, we show four examples of the reconstructed states at $B_z =$ 0.95, 1.05, 1.8, and 1.9, while $B_x$ is fixed to 0.1. 
In Fig.~\ref{FigS12}, we show the cumulative probability distribution (CPD) of the state fidelities, which gives a median of $97.4\%$.

For the experimental observation of Heisenberg scaling, we employ optimal measurements to achieve the theoretical maximum limit of CFI. 
We take the data points as a vector and quantify the total relative deviation with $\Vert\vec{\mathcal I}^\text{exp}(\mathcal M^\text{opt})-\vec{\mathcal I}^\text{sim}(\mathcal M^\text{opt})\Vert_2/\Vert\vec{\mathcal I}^\text{sim}(\mathcal M^\text{opt})\Vert_2$. This total relative deviation is $6.3\%$ for the experiments with different $B_x$ in Fig.3(b) in the main text. We attribute the discrepancy between the experimental and simulated Fisher information to the imperfect calibration of quantum gates, the decoherence and the sampling errors.

	\bibliographystyle{apsrev4-2}
	\bibliography{Supp_Reference}